\documentclass[copyright,creativecommons]{eptcs}
\providecommand{\event}{ICE 2017} 

\PassOptionsToPackage{hyphens}{url}
\usepackage{multicol}
\usepackage[english]{babel}
 \usepackage[utf8x]{inputenc}
\usepackage[usenames,dvipsnames]{xcolor} 
\usepackage{amsmath}
\usepackage{xargs}
\usepackage{amssymb}
\usepackage{times}
\usepackage{hyperref}
\usepackage[capitalise]{cleveref}
\usepackage{listings}
\usepackage{etoolbox}
\usepackage{comment}

\usepackage{graphicx}
\DeclareGraphicsExtensions{.pdf,.png,.jpg}
\graphicspath{{../pics}}
\usepackage{url}
\usepackage{stmaryrd}
\usepackage{array}
\usepackage{tikz}
\usetikzlibrary{arrows,shapes,automata,backgrounds,petri,positioning,fit,calc}
\usepackage[draft]{fixme}
\fxusetheme{color}
\FXRegisterAuthor{eM}{aeM}{\color{orange} {\underline{eM}}}
\FXRegisterAuthor{lg}{alg}{\color{blue} {\underline{LG}}}

\usepackage[T1]{fontenc}
\usepackage{nicefrac}
\usepackage{scalefnt}
\usepackage{textcomp} 

\usepackage{pdfsync}
\usepackage{mathtools}
\usepackage{slashed}

\mathtoolsset{showonlyrefs}
\usepackage{wrapfig}
\usepackage{relsize}

\usepackage{amsthm}
\usepackage{thmtools,thm-restate}
\usepackage{enumerate}
\usepackage{afterpage}

\newcommand{\ifempty}[3]{%
  \ifthenelse{\isempty{#1}}{#2}{#3}%
}

\usepackage{xspace}
\usepackage{listings}
\usepackage{textcomp}

\newcommand{\quo}[1]{\lq\lq {#1}\rq\rq}

\newcommand{\branch}{\&}

\newcommand{\rcdt}{\{l_i\colon T_i\}_{i\in I}}
\newcommand{\brancht}[1][\alpha]{\branch\rcdt}

\newcommand{\AT}[2]{#1\colon\! #2}

\newif\ifny\nytrue
\nytrue

\newif\ifvv\vvtrue
\vvfalse

\newif\ifkohei\koheitrue
\koheifalse

\newif\ifmarco\marcotrue
\marcofalse

\def\fps@figure{tp}      
\def\fps@table{tp}
\newcommand{\node}{\mathsf{n}}

\newcommand{\enewchan}[2]{\ensuremath{\mathtt{newChan}\ \AT{#1}{#2}}}

\newcommand{\NIL}{\mathbf{0}}

\newcommand{\MRED}[1][]{%
  \ensuremath{%
    \ifthenelse{\equal{#1}{}}{%
      \rightarrow\!\!\!\!\rightarrow%
    }{%
      \rightarrow\!\!\!\!\rightarrow_{#1}%
    }%
  }%
}

\newcommand{\RED}[1][]{%
  \ensuremath{%
    \ifthenelse{\equal{#1}{}}{%
      \longrightarrow%
    }{%
      \longrightarrow_{#1}%
    }%
  }%
}

\newcommand{\LLC}%
{{\mathsf{L}\mathsf{L}\mathsf{S}}^{\mathsf{C}}}

\newcommand{\LLA}%
{{\mathsf{L}\mathsf{L}\mathsf{S}}^{\mathsf{A}}}

\newcommand{\newlinchan}[3]
{\ensuremath{\mathtt{newCont}\ \AT{#1}{#2}\ \mathtt{in}\ #3}}

\newcommand{\newchan}[3]
{\enewchan{#1}{#2}\,;\,#3}

\newcommand{\SIL}%
{{\mathsf{S}\mathsf{I}\mathsf{L}}\xspace}

\newcommand{\SILc}%
{{\mathsf{S}\mathsf{I}\mathsf{L}^{\text{\scriptsize C}}}\xspace}

\newcommand{\SILC}%
{{\mathsf{S}\mathsf{I}\mathsf{L}^{\text{\scriptsize C}}}\xspace}

\newcommand{\SILa}%
{{\mathsf{S}\mathsf{I}\mathsf{L}^{\text{\scriptsize A}}}\xspace}

\newcommand{\SILA}%
{{\mathsf{S}\mathsf{I}\mathsf{L}^{\text{\scriptsize A}}}\xspace}

{\end{array}%
 \right.}

\newcommand{\causes}[1]{\searrow}

\newif\ifdm\dmtrue

\newif\ifdmr
\dmrfalse

\newcounter{analphabet}
{\rm%
\begin{list}%
{\arabic{analphabet}. }%
{\usecounter{analphabet}%
 \addtolength{\labelwidth}{5mm}%
\addtolength{\leftmargin}{-2mm}%
\setlength{\rightmargin}{0pt}%
\setlength{\itemsep}{0mm}%
\setlength{\parsep}{0pt}}}%
{\end{list}}

\DeclareSymbolFont{bbsymbol}{U}{bbold}{m}{n}
\DeclareMathSymbol{\bbsemicolon}{\mathbin}{bbsymbol}{"3B}

\newcommand{\MSACore}%
{CMSA\xspace}
\newcommand{\MSA}%
{MSA\xspace}

\newcommand{\eMhide}[1]{}

\newcommand{\toolidcol}{black}

\newcommand{\mycode}[1]{
\begin{verbatim}{#1}\end{verbatim}
}

\newcommand{\toolid}[1]{\textcolor{\toolidcol}{{\small\textsf{#1}}}}
\newcommand{\chorgram}{\toolid{ChorGram}}

\def\colorPtp{\color{Blue}}

\def\colorNode{\color{cyan}}
\def\colorFun{\color{red}}
\def\colorR{\color{OliveGreen}}
\def\colorE{\color{orange}}

\definecolor{light-gray}{gray}{0.97}

\makeatletter
\newcommand{\linenumfontsize}{\@setfontsize{\linenumfontsize}{3pt}{3pt}}
\makeatother
\lstdefinelanguage{sys}{
	commentstyle=\color{Gray},
	morecomment=[s]{[}{]},
	keywords=[0]{system,of,do,end},	keywordstyle=\color{orange}\bfseries,
}

\newcommand{\gatedistancein}{3pt}
\newcommand{\gatedistanceinand}{2pt}

\newcommand{\msg}[1][m]{\mathsf{#1}}

\newcommand{\ptp}[1][A]{\ensuremath{\mathsf{\colorPtp{\MakeUppercase{#1}}}}}

\newcommand{\p}{\ptp}
\newcommand{\q}{{\ptp[B]}}

\newcommandx{\common}[3][1=\ptp,2={\aR},3={\aR'},usedefault=@]{f_{#1}}
\newcommandx{\opair}[2][1={\ae},2={\ae'},usedefault=@]{\langle {#1},{#2} \rangle}
\newcommandx{\hopair}[2][1={\aE},2={\aE'},usedefault=@]{\llparenthesis\, {#1},{#2}\, \rrparenthesis}

\newcommandx{\wb}[2][1={\aG},2={\aG'},usedefault=@]{wb({#1}, {#2})}
\newcommandx{\ws}[2][1={\aseq},2={\aseq'},usedefault=@]{ws({#1}, {#2})}
\newcommandx{\widx}[2][1={\aW},2={i},usedefault=@]{{#1}[{#2}]}
\newcommandx{\outop}[2][1=\gname,2={}]{!^{{#1}{#2}}}
\newcommandx{\inop}[2][1=\gname,2={}]{?^{{#1}{#2}}}
\newcommandx{\aout}[5][1={\p},2={\q},3=\gname,4=\msg,5={},usedefault=@]{
  \achan[{#1}][{#2}] \outop[{#3}] {#4}{#5}
}
\newcommandx{\ain}[5][1={\p},2={\q},3=\gname,4=\msg,5={},usedefault=@]{
  \achan[{#1}][{#2}] \inop[{#3}] {#4}{#5}
}
\newcommandx{\adep}[1][1={}]{
  \langle \aout[@][@][@][@][{#1}], \ain[@][@][@][@][{#1}] \rangle
}

\newcommandx{\hproj}[2][1=\aR, 2=\ptp, usedefault=@]{
  \ifempty{#1}{}{{#1}}\ifempty{#2}{}{{^{\scriptscriptstyle @{#2}}}}
}
\newcommandx{\eproj}[2][1=\aE,2=\ptp, usedefault=@]{
  {{#1}}\ifempty{#2}{}{{^{\scriptscriptstyle @{#2}}}}
}

\newcommandx{\fsaout}[2][1={\p},2={},usedefault=@]{
  \ptp[{#1}] \ \outop[]\ \msg[{#2}]
}
\newcommandx{\fsain}[2][1={\p},2={},usedefault=@]{
  \ptp[{#1}] \ \inop[]\ \msg[{#2}]
}

\newcommand{\aM}{M}
\newcommandx{\cm}[2][1=\ptp, 2=\aM]{{#2}_{#1}}
\newcommandx{\achan}[2][1=A,2=B,usedefault=@]{\msg[#1]\cdot\msg[#2]}

\newcommand{\conf}[1]{\langle #1 \rangle}

\newcommand{\aG}{\mathsf{G}}

\newcommand{\gseqop}{;}
\newcommand{\gparop}{\,|\,}
\newcommand{\gchoop}{+}
\newcommand{\grecop}{*}
\newcommand{\grecopp}{@}
\newcommand{\gname}[1][i]{{\colorNode{\scriptstyle\textsf{#1}}}}

\newcommandx{\gnode}[2][1=\gname,2=\gint,usedefault=@]{
  {\ifempty{#1}{}{\colorNode{#1.}}} {#2}
}

\newcommandx{\gint}[4][1=\gname,2=\ptp,3=\msg,4={\ptp[B]},usedefault=@]{
  \gnode[{#1}][{#2} \xrightarrow{} {#4} \colon {\msg[{#3}]}]
}
\newcommandx{\gseq}[3][1=\gname,2={\aG},3={\aG'},usedefault=@]{
  \gnode[{#1}][{#2} \gseqop {#3}]
}
\newcommandx{\gpar}[3][1=\gname,2={\aG},3={\aG'},usedefault=@]{
  \gnode[{#1}][\ifempty{#1}{{#2} \gparop {#3}}{({#2} \gparop {#3})}]}
\newcommandx{\gcho}[3][1=\gname,2={\aG},3={\aG'},usedefault=@]{
  \gnode[{#1}][\ifempty{#1}{{#2} \gchoop {#3}}{({#2} \gchoop {#3})}]
}
\newcommandx{\grec}[2][1={\aG},2={\p},usedefault=@]{
  \grecop {#1} \grecopp {#2}
}

\newcommandx{\gsem}[2][1={\aG},2={},usedefault=@]{[\![ {#1} ]\!]_{#2}}
\newcommandx{\rbot}{\text{undef}}
\newcommandx{\rtrs}[1][1={\aR},usedefault=@]{{#1}^{\star}}
\newcommandx{\gord}[1][1={\aG},usedefault=@]{<_{#1}}
\newcommandx{\gordeq}[1][1={\aG},usedefault=@]{\leq_{#1}}

\newcommandx{\rlang}{\mathcal{L}}
\newcommandx{\efst}[1]{\textsf{cs}\ifempty{#1}{}{_{({#1})}}}

\newcommandx{\aW}{w}

\newcommand{\gfun}[1]{\ensuremath{\mathsf{\colorFun #1}}}

\newcommandx{\rseq}[2][1=\aG,2={\aG'},usedefault=@]{\gfun{seq}({#1},{#2})}
\newcommandx{\rpar}[2][1=\aG,2={\aG'},usedefault=@]{\gfun{par}({#1},{#2})}

\newcommand{\aR}[1][R]{{\colorR{#1}}}
\renewcommand{\ae}[1][e]{{\colorE{#1}}}

\newcommandx{\gproj}[2][1=\aG,2=\ptp]{{#1}\downarrow_{#2}}
\newcommandx{\cinit}[1][1={q_0},usedefault=@]{{#1}}
\newcommandx{\cfinal}[1][1={q_e},usedefault=@]{{#1}}

\newcommandx{\geproj}[4][1=\aG,2=\ptp,3=\cinit,4=\cfinal,usedefault=@]{
  {#1}\downarrow_{#2}^{{#3},{#4}}
}

\newcommand*{\StrikeThruDistance}{0.15cm}%
\tikzset{strike thru arrow/.style={
    decoration={markings, mark=at position 0.5 with {
        \draw [blue, thick,-] 
            ++ (-\StrikeThruDistance,-\StrikeThruDistance) 
            -- ( \StrikeThruDistance, \StrikeThruDistance);}
    },
    postaction={decorate},
}}

\tikzset{smallglobal/.style={
  node distance=1cm and 0.8cm, semithick, scale=0.8, every node/.style={transform shape}
}}

\newcommand{\sourceG}{
  \begin{tikzpicture}[smallglobal,baseline=-.5ex, scale=0.6, every node/.style={transform shape}]
  \node [source] (src) {};
\end{tikzpicture}
}

\newcommand{\sinkG}{
\begin{tikzpicture}[smallglobal,baseline=-.5ex, scale=0.6, every node/.style={transform shape}]
  \node [sink] (src) {};
\end{tikzpicture}
}

\newcommand{\orgateG}{
\begin{tikzpicture}[smallglobal,baseline=-.5ex, scale=0.75, every node/.style={transform shape}]
  \node [ogate] (o) {};
\end{tikzpicture}
}

\newcommand{\andgateG}{
\begin{tikzpicture}[smallglobal,baseline=-.5ex, scale=0.75, every node/.style={transform shape}]
  \node [agate] (o) {};
\end{tikzpicture}
}

\newcommand{\bpmnchoice}{
\begin{tikzpicture}[smallglobal,baseline=-.5ex, scale=0.75, every node/.style={transform shape}]
  \node [bpmnchoice] (o) {$\times$};
\end{tikzpicture}
}

\newcommand{\bpmnchoiceempty}{
\begin{tikzpicture}[smallglobal,baseline=-.5ex, scale=0.75, every node/.style={transform shape}]
  \node [bpmnchoice] (o) {\color{white}{$\times$}};
\end{tikzpicture}
}

\newcommand{\bpmnpar}{
\begin{tikzpicture}[smallglobal,baseline=-.5ex, scale=0.75, every node/.style={transform shape}]
  \node [bpmnchoice] (o) {$+$};
\end{tikzpicture}
}

\newcommandx{\ich}[1][1={\aG},usedefault=@]{{#1}^{\oplus}}
\newcommandx{\ichedges}[2][1={\aG},2={\gname},usedefault=@]{{#1}^{\oplus}({#2})}
\newcommandx{\parts}[1]{2^{#1}}
\newcommandx{\actch}{c}
\newcommandx{\soundactch}[2][1={\aG},2={\actch},usedefault=@]{{#1} \,\circledR\, {#2}}
\newcommandx{\rOnActch}[2][1={\aG},2={\actch},usedefault=@]{{#1} \setminus {#2}}
\newcommandx{\rOnActchClean}[2][1={\aG},2={\actch},usedefault=@]{{#1} \circledR {#2}}
\newcommandx{\rAllEvents}[1][1={\aG},usedefault=@]{\mathit{dom}(#1)}
\newcommand{\aE}{{\tilde \ae}}

\newcommandx{\hyedge}[1]{\{#1\}}

\newcommandx{\rdiv}[2][1=\gcho,2=\ptp,usedefault=@]{
  \gfun{div}_{#2}(#1)
}

\newcommandx{\rrdiv}[5][1={\aG},2={\aG'},3={\aE},4={\aE'},5=\ptp,usedefault=@]{
  \gfun{div}^{#3,#4}_{#5}(#1,#2)
}
\newcommandx{\pdiv}[3][1={\apom_1},2={\apom_2},3={\apom},usedefault=@]{
  \gfun{div}_{#3}(#1,#2)
}
\newcommandx{\pfork}[3][1={\apom_1},2={\apom_2},3={\apom},usedefault=@]{
  \gfun{fork}_{#3}(#1,#2)
}

\newcommandx{\aQzero}[1][1=,usedefault=@]{
  {\ifempty{#1}{q_0}{q_{q#1}}}
}

\tikzset{
  src/.style={draw,circle,fill=white,
    minimum size=2mm,
    inner sep=0pt},
  sink/.style={draw,circle,double,fill=white,
    minimum size=1.5mm,
    inner sep=0pt},
  node/.style={draw,circle,fill=black,
    minimum size=2mm,
    inner sep=0pt},
source/.style={draw,circle,fill=white,
  minimum size=4mm,
  inner sep=0pt},
sink/.style={draw,circle,double,fill=white,
  minimum size=3mm,
  inner sep=0pt},
  block/.style = {rectangle, draw=gray, align=center, fill=orange!25, rounded corners=0.1cm,
    minimum size=5mm, inner sep=2pt},
  prenode/.style = {minimum size=9pt,inner sep=2pt, font=\Large},
  bblock/.style = {rectangle, draw=blue!50, opacity=.5, line width=1pt, align=center, fill=white, rounded corners=0.1cm,
    minimum size=7mm, inner sep=2pt},
  prenode/.style = {minimum size=9pt,inner sep=2pt, font=\Large},
  altgate/.style={draw, rectangle,
    minimum size=3mm,
    inner sep=0pt,
    postaction={path picture={%
        \draw
        ([yshift=\gatedistanceinand]path picture bounding box.south) --
        ([yshift=-\gatedistanceinand]path picture bounding box.north) ;}}},
  agate/.style={draw, rectangle,
    minimum size=3mm,
    inner sep=0pt,
    postaction={path picture={%
        \draw
        ([yshift=\gatedistanceinand]path picture bounding box.south) --
        ([yshift=-\gatedistanceinand]path picture bounding box.north) ;}}},
  ogate/.style = {
    diamond, draw,
    minimum size=4mm,
    inner sep=0pt,
    postaction={path picture={%
        \draw
        ([yshift=\gatedistancein]path picture bounding box.south) -- ([yshift=-\gatedistancein]path picture bounding box.north)
        ([xshift=-\gatedistancein]path picture bounding box.east) -- ([xshift=\gatedistancein]path picture bounding box.west)
        ;}}},
  altogate/.style = {
    diamond, draw,
    minimum size=4mm,
    inner sep=0pt,
    postaction={path picture={%
        \draw
        ([yshift=\gatedistancein]path picture bounding box.south) -- ([yshift=-\gatedistancein]path picture bounding box.north)
        ([xshift=-\gatedistancein]path picture bounding box.east) -- ([xshift=\gatedistancein]path picture bounding box.west)
        ;}}},
  anygate/.style = {circle, draw, fill=white,
    minimum size=4mm,
    inner sep=0pt,
    postaction={path picture={%
        \draw[black]
        ([xshift=-\gatedistancein,yshift=\gatedistancein]path picture bounding box.south east) --
        ([xshift=\gatedistancein,yshift=-\gatedistancein]path picture bounding box.north west)
        ([xshift=-\gatedistancein,yshift=-\gatedistancein]path picture bounding box.north east) --
        ([xshift=\gatedistancein,yshift=\gatedistancein]path picture bounding box.south west)
        ;}}},
  elli/.style = {draw,densely dotted,-},
  line/.style = {draw,->, rounded corners=0.07cm,>=latex},
  nline/.style = {draw,semithick, ->},
  pline/.style = {draw,->,>=latex},
  node distance=1cm and 0.7cm,
  baseline=(current  bounding  box.center),
  bpmnchoice/.style = {
    diamond, draw,
    minimum size=4mm,
    inner sep=0pt,
  },
}

\newcommand{\hidden}[1]{}

\usepackage{mymacros}

\newtheorem{fact}{Fact}[section]

\newtheorem{definition}[fact]{Definition}

\newcommand{\sysname}{ACC}

\newcommand{\IoTLySa}{{\toolid{IoT-LySa}}}

\newcommand{\mess}[1]{\mbox{\guillemotleft}{#1}\mbox{\guillemotright}}

\lstnewenvironment{lysacode}{\lstset{language=C,
		extendedchars=true,
		showstringspaces=false,
                escapechar=£,
                commentstyle=\color{gray},
		basicstyle=\sffamily\scriptsize,
		morekeywords={actuator, then, snd, recv, rec, nil, func, bool, 
			sensor, process, actuator_type, node,type,to},
}}{}

\title{Tool Supported Analysis of IoT}
\author{Chiara Bodei \qquad Pierpaolo Degano \qquad Letterio Galletta
\institute{Dipartimento di Informatica, Universit\`a di Pisa, \\ Pisa, Italy}
\email{ \{chiara,degano,galletta\}@di.unipi.it}
\and
Emilio Tuosto
\institute{Department of Informatics, University of Leicester, \\ Leicester, United Kingdom}
\email{emilio@leicester.ac.uk}
}
\def\titlerunning{Tool Supported Analysis of IoT}
\def\authorrunning{C.~Bodei, P.~Degano, L.~Galletta \& E.~Tuosto}
\begin{document}
\maketitle

\begin{abstract}
  The design of IoT systems could benefit from the combination of two
  different analyses.
  We perform a first analysis to approximate how data flow across the
  system components, while the second analysis checks their
  communication soundness.
  We show how the combination of these two analyses yields further
  benefits hardly achievable by separately using each of them.
  We exploit two independently developed tools for the analyses.
  
  Firstly, we specify IoT systems in \IoTLySa, a simple specification language 
  featuring asynchronous multicast communication of tuples.
  The values carried by the tuples are drawn from a term-algebra
  obtained by a parametric signature.
  The analysis of communication soundness is supported by \chorgram, a
  tool developed to verify the compatibility of communicating
  finite-state machines.
  In order to combine the analyses we implement an encoding of 
  \IoTLySa\ processes into communicating machines.
  This encoding is not completely straightforward because \IoTLySa\ has
  multicast communications with data,  while
  communication machines are based on point-to-point communications
  where only finitely many symbols can be exchanged.
  To highlight the benefits of our approach we appeal to a simple yet
  illustrative example.
\end{abstract}

\section{Introduction}\label{sec:intro}
%
Software is hardly built anymore as a stand-alone application.
Over the last decades, the rise of distributed and concurrent
applications has determined a crucial entanglement between computation
and communication~\cite{mil93turing}.
This phenomenon increasingly makes adaptability, reconfigurability,
and openness paramount requirements of software.
%
%
Therefore, it is nowadays crucial for designers and software
engineers to take into account both properties of computation
and communication when developing systems.

The Internet of Things (IoT) is one of the most advanced technological
frontiers of the combination of computation and communication.
By definition, IoT systems are made of several components that
interact within complex (cyber-)physical environments.
This emerging technology provides the momentum for the creation of new
societal and economical opportunities, as long as IoT systems
become trustworthy.
Indeed, the great disruptive innovation that IoT systems can offer
comes with an equally great threat: it is extremely hard to guarantee
correctness properties of those systems.
In fact, the overall correctness becomes the resultant of that
of communication and that of computation.
\mbox{In Bruce Schneier's words:}
\begin{quote}
  \guillemotleft{\it The Internet of Things is fundamentally changing how computers get
  incorporated into our lives. Through the sensors, we're giving the
  Internet eyes and ears. Through the actuators, we're giving the
  Internet hands and feet. Through the processing -- mostly in the
  cloud -- we're giving the Internet a brain. Together, we're creating
  an Internet that senses, thinks, and acts. This is the classic
  definition of a robot, and I contend that we're building a
  world-sized robot without even realizing it.}\guillemotright
  \ \cite{sch17}
\end{quote}

%
%
%
These issues are particularly critical because of the apparently easy use of
IoT technologies hides their not trivial design, that becomes evident
only when something goes wrong, e.g.~when privacy can be compromised.
Among many others, a recent example is that of a smart
doll~\footnote{%
  See e.g.\
  \url{https://www.theguardian.com/world/2017/feb/17/german-parents-told-to-destroy-}
  \url{my-friend-cayla-doll-spy-on-children}
} 
that allows children to access the Internet via
speech recognition software, and to control the toy via an app.
However, this smart technology enables an attacker to also access personal data.

%

In this paper, we consider a simple yet realistic scenario to
illustrate how to verify properties of IoT systems that arise from
this prominent mix of computation and communication.
We advocate the combination of two different techniques for supporting verification.
To analyse the \quo{computational dimension} we appeal to a static
analysis technique that allows us to check control and data flow
properties of systems.
The soundness of \quo{communication dimension} is instead verified
through a choreographic framework.
More precisely, we combine the analysis defined on
\IoTLySa~\cite{BDFG_Coord16,BDFG_ICE2016} (a specification language
tailored to IoT) and the analysis of communication soundness based on
communicating finite-state machines (CFSMs) and global
graphs~\cite{lty15}.

An IoT system is specified as a set of \IoTLySa\ components that
communicate through an asynchronous multicast mechanism.
The values communicated in the system are drawn from a term-algebra
obtained by a parametric signature.
The control flow analysis for \IoTLySa~\cite{BDFG_Coord16} over-approximates
the values that can be assigned to the local variables of
a component, as well as where these values originated.
This analysis allows designers to spot potential mistakes on how data
are handled by components.
In addition, the analysis computes the messages that a node may receive, but not the order in which they arrive, and it is thus insufficient for establishing the communication soundness.

Communication soundness is therefore checked by mapping the components to CFSMs
and using \chorgram~\cite{chorgram}, a tool developed to verify the \emph{compatibility} of
communicating finite-state machines~\cite{lty15}.
The compatibility property checked by \chorgram\ is defined
in~\cite{lty15} and it is a condition guaranteeing absence of
deadlocks, message obliviousness (i.e.\ messages sent but not
received)~\cite{bstz16}, and of unspecified reception (i.e.\ a
component receiving unexpected messages).
We show how \chorgram\ flags possible communication mistakes when the
set of CFSMs is not compatible and allows the designer to determine
where the problem arises.
Differently from the control flow analysis, the one made with \chorgram\ is insensitive to the data exchanged through the messages.

The main contribution of this paper is showing that the control flow analysis and the analysis of communication soundness based on a choreographic framework complement each other, and that it is worthwhile combining them for formally verifying IoT systems.

\paragraph{Structure of the paper}
After reviewing the choreographic framework and \IoTLySa\
(cf.\ Section~\ref{sec:bkg}), we describe our running example and model
it in \IoTLySa\ (cf.\ Section~\ref{sec:rt}).
Section~\ref{sec:rt} also shows how the analysis of the \IoTLySa\
specification can lead to detecting mistakes in the computation
and help in fixing the errors.
Section~\ref{sec:comm} illustrates how to use \chorgram\ to
analyse the communication behaviour of our running example.
This analysis flags some unexpected misbehaviour and helps 
fixing the problem.
In order to combine these analyses Section~\ref{sec:encoding} describes
an encoding of the \IoTLySa\ systems into communicating machines.
This encoding is not straightforward because \IoTLySa\
features multicast communications with data while communication
machines are based on point-to-point communications where only
finitely many symbols can be exchanged.
Technical details  are given in the appendices.
Our concluding remarks are in Section~\ref{sec:conc}.


\section{Background}\label{sec:bkg}
\subsection{A choreographic framework for distributed systems}\label{sec:cho}

%
Choreographies advocate the so-called \emph{local} and \emph{global} views
for specifying the behaviour of a system.
In the first, one specifies the asynchronous communication behaviour
of each component \quo{in isolation}, while the interactions among the
system components are seen as atomic actions that abstract away from
asynchrony.
%
%
Using a simple example, we will review the key ingredients of the
local and global views adopted here, namely communicating finite-state
machines~\cite{bz83} and (a variant of) global graphs~\cite{dy12}.

\paragraph*{Local view} Communicating finite-state machines (CFSMs)
are a conceptually simple model for the analysis of communication
protocols.
A CFSM is basically a finite-state automaton with transitions that
represent send or receive actions.
Each machine is uniquely identified by a name and we use
$\p$, $\q$, etc. to range over identifiers of CFSMs.
A system consists of finitely many CFSMs that share unbounded FIFO
channels to exchange messages.
More precisely, for each pair $\p$ and $\q$ of CFSMs there is a
channel from $\p$ to $\q$ (dubbed $\achan$) and a channel in the other
direction (dubbed $\achan[\q][\p]$).
The use of buffers realises an asynchronous semantics: senders simply
deposit messages in buffers without synchronising with receivers.
\begin{figure}[t]
  \centering
  \begin{tabular}{c@{\qquad\qquad}c}
    \includegraphics[scale=.4]{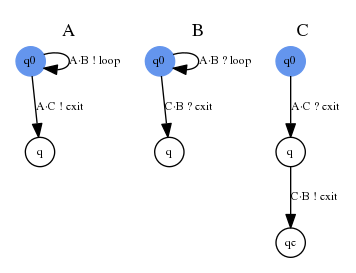}
    &
      \includegraphics[scale=.25]{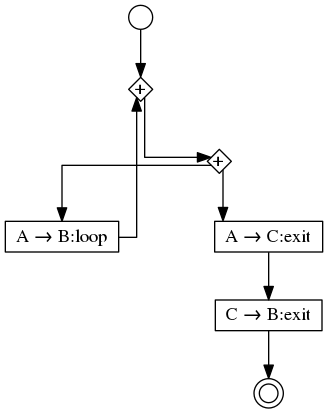}
    \\
    (a) Example of a system of CFSMs & (b) Global graph
  \end{tabular}
  \caption{A choreographic model}
  \label{fig:chor}
\end{figure}
We illustrate the semantics of CFSMs by considering the system of
three machines $\p$, $\q$, and $\ptp[C]$ in
\figurename~\ref{fig:chor}(a).
Machine $\p$ repeatedly sends message $\msg[loop]$ to machine $\q$
until message $\msg[exit]$ is sent to machine $\ptp[C]$.
Machine $\q$ will consume any message sent by $\p$ and then wait for a
message $\msg[exit]$ in the buffer $\achan[{\ptp[C]}][\q]$ from
$\ptp[C]$.
Finally, $\ptp[C]$ forwards message $\msg[exit]$ to $\q$ once
it has received it from $\p$.
The \emph{configuration} of a system is a snapshot that records the
states in which each machine is in, and the contents of each channel.
Initially, all CFSMs are in their initial states and buffers are empty.
When, in the current configuration, a machine is in a state with an
outgoing output transition, labelled say by $\aout[@][@][]$, then
the output transition can be fired and move the machine $\p$ to the target state of
the transition.
At the same time the message $\msg$ is deposited in the buffer $\achan$.
Similarly, if the machine is in a state with an outgoing input
transition, labelled say by $\ain[@][@][]$, and the head
of the buffer contains message $\msg$ then the input transition
can be fired removing $\msg$ from the buffer and moving the
machine to the target state of the transition.
For instance, the system of our example in
\figurename~\ref{fig:chor}(a) can reach the following configuration
 (from its initial one)
\[
  \conf{\p:q,\q:q_0,\ptp[C]:q_0  \ \ ; \ \ \achan:[\msg[loop],\ldots,\msg[loop]], \achan[@][{\ptp[C]}] : [\msg[exit]], \achan[{\ptp[C]}][@]:[],\ldots}
\]
where (i) $\p$ is in state $q$ while machines $\q$ and $\ptp[C]$ are
in state $q_0$ and (ii) the buffer $\achan$ contains a number of
$\msg[loop]$ messages, the buffer $\achan[@][{\ptp[C]}]$ contains
the message $\msg[exit]$, and all the remaining buffers are empty.
For simplicity, suppose that $\p$ sends only one $\msg[loop]$ message
and then an $\msg[exit]$ message.
In this configuration, both $\q$ and $\ptp[C]$ can consume the message
in their queues from $\p$; if the input of $\ptp[C]$ is fired then the
system reaches the configuration
\[
  \conf{\p:q,\q:q_0,\ptp[C]:q  \ \ ; \ \ \achan:[\msg[loop]], \achan[@][{\ptp[C]}] : [], \achan[{\ptp[C]}][@]:[],\ldots}
\]
from where both the input of $\msg[loop]$ of $\q$ and the output from
state $q$ of $\ptp[C]$ are enabled and can be non-deterministically
fired.
Observe that the computation where $\p$ never sends message
$\msg[exit]$ is possible; this will make $\p$ and $\q$ interact
forever, while $\ptp[C]$ will be waiting.

\paragraph*{Global view}
A global graph yields the description of the distributed work-flow of
components exchanging messages.
We borrow global graphs from~\cite{lty15} to formalise the global
views of choreographies; an example of these graphs is in
\figurename~\ref{fig:chor}(b).
The nodes of a global graph are labelled\footnote{
  The notation of the global graphs used in~\cite{lty15} is
  reminiscent of BPMN choreography~\cite{bpmn}.
  The reader familiar with BPMN should note that our \orgateG-node
  corresponds to the \bpmnchoice and \bpmnchoiceempty gateways in
  BPMN, while our \andgateG-node corresponds to the \bpmnpar gateway
  in BPMN.  } 
  by  \sourceG (resp \sinkG) to mark the starting
(resp. termination) point of the choreography;
by $\p \to \q : \msg$
to indicate that participant $\p$ and $\q$ exchange message $\msg$
($\p$ being the sender and $\q$ the receivers);
by \orgateG to
indicate either a choice, merge, or the entry-point of an iteration;
or by \andgateG to indicate either the start or the end of concurrent
interactions.
For example, the starting point of the global graph in~\figurename~\ref{fig:chor}(b) 
is followed by the entry point of an
iteration; the body of the iteration is a choice between two branches:
in the left-most branch $\p$ and $\q$ exchange the message
$\msg[loop]$ and go back to the entry point of the iteration; in the
right-most branch $\p$ and $\ptp[C]$ exchange the message $\msg[exit]$
followed by an exchange of the same message between $\ptp[C]$ and
$\q$, followed by a sink node where no further interactions happen.

An intuition of the semantics of a global graph can be given in terms
of a \quo{token game} similar to the one of Petri nets.
The token starts from the initial node and follows the arcs of the
graph; when the node reaches a choice node \orgateG it continues along
one branch; when the node reaches a fork node
\andgateG, new tokens start to flow along all the arcs and, dually, when
all the tokens of the incoming arcs of a join gate are present then
one token continues along the outgoing arc of the gate.
For example, in our global graph, after the start, the machines either
engage in an iteration where $\p$ and $\q$ exchange $\msg[loop]$ or $\p$
sends the message $\msg[exit]$ to $\ptp[C]$, which in turn forwards it
to $\q$ after which the protocol terminates.

In this paper, we will use \chorgram ~\cite{chorgram} a toolkit, based
on the results of~\cite{lty15}, that allows to reconstruct
a global graph out of a system of CFSMs.
In fact,  \chorgram \ generates the  presented global graph starting from the
machines $\p$, $\q$, and $\ptp[C]$ above.

As said above, systems of CFSMs evolve by moving from one
\emph{configuration} to another.
If communication misbehaviours happen then \quo{bad} configurations
may be reached.
A machine is in a \emph{sending} (resp.\ \emph{receiving}) state if
all its outgoing transitions are send (resp.\ receive) actions.
A state without any outgoing transition is said to be \emph{final}.
A state that is neither final, nor sending nor receiving is a \emph{mixed
  state}.
A system of CFSMs is \emph{communication sound} if none of its
reachable configurations is a \emph{deadlock} and it is not
\emph{message oblivious}~\cite{bstz16}, where
\begin{itemize}
\item a configuration is a \emph{deadlock} if all the buffers are
  empty, there is at least one machine in a receiving state, and all
  the other machines are either in a receiving state or in a final
  state;
\item a system is not \emph{message oblivious}~\cite{bstz16} if
  whenever a message has been sent by a participant, that message will
  be eventually received.\footnote{Like eventual
    reception~\cite{BLY15}, message obliviousness is stronger than the
    no-orphan message property; a non message oblivious system is also
    free of orphan messages.}
\end{itemize}
Since CFSMs are Turing-complete, communication soundness is not
decidable.
However, systems enjoying the \emph{generalised multiparty
 compatibility} (GMC)~\cite{lty15} are guaranteed to be sound and GMC
is a decidable property supported by the \chorgram\
toolkit~\cite{lty17}.
We will give an intuition of the GMC property in Section~\ref{sec:comm}.


%
\subsection{An overview of \IoTLySa}\label{sec:iotlysa}
%

\paragraph*{The language}
We review \IoTLySa~\cite{BDFG_Coord16,BDFG_ICE2016}, a specification
language recently proposed for designing IoT systems.
\IoTLySa\ is based on a process calculus and it has a specific focus
on data collecting, aggregation, and communication.
%
%
%
Indeed, by using \IoTLySa\ primitives designers can model the
structure of a system and the interactions among the smart objects it is made of.
A smart object is represented by a node that
specifies multiple processes interacting through a shared
local store, as well as sensors and actuators.
In
\IoTLySa \ store accesses are assumed as atomic.
The processes are in charge of detailing how data are to be processed
and exchanged among the nodes.
A sensor of a node $\ell$ is an active entity that reads data from the physical
environment at its own fixed rate, and deposits them in the local
store of $\ell$.
Actuators instead are passive: they just wait for a command to become
active and operate on the environment.
The current version of \IoTLySa\ is specifically designed to model
monitoring system typical of smart cities, factories or farms.
In this scenario, smart objects never leave their locations, while
mobile entities, such as cars or people, carry no smart device and can
only trigger sensors.

Communications in IoT systems often occur in a disciplined broadcasting manner.
For that, \IoTLySa\ features a simple multicast communication among nodes.
For example, the output of the first thermometer node \lstinline!Th$_1$! at line~\ref{line:a-entrambi} in \figurename~\ref{fig:versione1}
\begin{center}
\lstinline!snd (temp,mt) to [Th,Th$_2$]!
\end{center}
specifies that the message \lstinline!(temp,mt)! has to be sent to two nodes, namely the thermostat
\lstinline!Th! and the second thermometer \lstinline!Th$_2$! (we will come back on this example in Section~\ref{sec:rt}).
The reception of a message is also done in a disciplined way: the receiver filters the messages flowing in the ether and only accepts some of them according to a sort of guard, following~\cite{BBDNN_JCS}.
For example, consider the input of the thermostat node \lstinline!Th! in
\figurename~\ref{fig:nodo-semplice} at line~\ref{line:rcvcont}
\begin{center}
\lstinline!recv(temp;x)!
\end{center}
The message described above passes the check, because what is at the left of the ``\lstinline!;!'' in the payload, namely the guard \lstinline!temp!, matches with the first element of the message sent.
Then the value stored in \lstinline!mt! is assigned to the variable \lstinline!x!, as usual.
If the guard is not passed, \lstinline!Th!  simply ignores the message.
(Of course, the above extends to tuples of guards and of binding values.)
This filtering input enables us to explicitly implement external
choices, through the \lstinline!switch! primitive.  It was not
originally included in~\cite{BDFG_Coord16,BDFG_ICE2016} although the construct can be
encoded there.

We now intuitively describe other elements of \IoTLySa\ on the snippet of code in
\figurename~\ref{fig:nodo-semplice}.
As done already, for readability, in this paper we are using a sugared
syntax, from which the original \IoTLySa\ code can be easily
extracted (see Appendix~\ref{sec:applysa}).
The thermostat node \lstinline+Th+ we are considering has no sensors and a single
actuator \lstinline+ACon_off+.
It is part of the system further discussed in Section~\ref{sec:rt} that keeps the temperature of a room under
control with an air conditioning, and that includes two further nodes, describing two thermometers.
\begin{figure}

\begin{lysacode}
node Th = 
     actuator ACon_off : onoff_t =  rec h. wait_for(turnon, turnoff); h £\label{line:aconoff}£

     process = 
         rec h. £\label{line:rec}£
            switch {
              recv(temp;x); £\label{line:rcvcont}£
              if tthreshold < x then £\label{line:guard}£
                  @ACon_off.turnon;
                  snd (ack,on) to [Th$_1$]; £\label{line:sndon}£
                  h
              else
                  @ACon_off.turnoff;
                  snd (ack,off) to [Th$_1$]; £\label{line:else}£
                  h
            |
              recv(quit;i); £\label{line:rcvquit}£
              nil
            } £\label{line:end}£
\end{lysacode}
\caption{The specification of a thermostat \lstinline+Th+ in \IoTLySa}
\label{fig:nodo-semplice}
\end{figure}

The behaviour of the actuator \lstinline!ACon_off! (line~\ref{line:aconoff}) depends on
commands \lstinline+turnon+ and \lstinline+turnoff+ to 
start and stop the air conditioning, respectively.
(Note that the actual behaviour is not described, as the physical environment is seen as a black box.)
The behaviour of the node \lstinline+Th+ is an iterative
process (lines~\ref{line:rec}-\ref{line:end}).
The loop consists of an external choice rendered with the
\lstinline+switch+ command that declares two branches guarded by
two mutually exclusive inputs \lstinline+recv+.
More precisely, the first branch is executed when the node receives
a message with guard \lstinline+temp+ (line~\ref{line:rcvcont}).
The node checks if the temperature
\lstinline+x+ is below or above
a given threshold, instructs the actuator, and sends an
acknowledgement accordingly (lines~\ref{line:guard}-\ref{line:else}).
The second case is when the node receives
the message with guard \lstinline+quit+ (line~\ref{line:rcvquit}):
the process sends an acknowledgement and
terminates.
\paragraph*{The control flow analysis}
%
A control flow analysis (CFA) that safely approximates the behaviour
of \IoTLySa\ systems has been given in~\cite{BDFG_Coord16}.
The analysis tracks how sensor values are used inside their local
nodes, how they are aggregated and propagated to other nodes, and
which messages are exchanged among nodes.
%
As any other static analysis it mimics the evolution of a system, collecting the relevant information.
In this case it traverses the text of the system and records the
effects of each possible aggregation function application, of the
assignments on the variables used in the specification, as well as
those of the possible inter-node communications.
As expected, the analysis uses abstract values in
place of concrete ones.
%

Roughly speaking, the abstract values of the data generated by an IoT system
describe their provenance and how they are
aggregated during the computations.
A natural way of abstracting concrete data would then be through trees, 
the leaves of which either represent sensor names (to identify the
sensors from which values come) or other \quo{primitive}
(abstract) values, e.g.\ constants.
The aggregation functions then combine available trees to obtain new ones.
However, IoT systems typically have feedback loops (as in our running
example), which are realised by some iterative behaviour.
This may prevent us from statically determining safe approximants of aggregated
data, because we cannot establish how many
applications of aggregation functions are necessary by only inspecting the code.
The analysis of~\cite{BodeiDFG16a} resorts therefore to tree grammars
for keeping finite the abstract values and for improving the precision of the analysis
in~\cite{BDFG_Coord16}, where instead trees are cut at a fixed depth.

We will come back on the above rather technical issue in Section~\ref{sec:rt}, and here we only intuitively illustrate the results of the analysis and some of its possible uses.
For each node $\ell$ in the network, the analysis returns: 
\begin{itemize}
\item a map $\widehat{\Sigma}_\ell$ assigning to each sensor and each
  variable a set of the abstract values (safely representing the actual values) that they may assume at run
  time (recall that each node has its own local store);
\item a set $\kappa(\ell)$ that over-approximates the set of messages
  received by the node $\ell$, and for each of them its sender (basically,
  an element of $\kappa(\ell)$ is a pair sender-tuple of abstract
  values);
\item a set $\Theta(\ell)$ of possible abstract values, i.e.\ tree grammars, computed and
  used by the node $\ell$.
\end{itemize}
\noindent
The components $\kappa$ and $\Theta$ track how data may flow in the
network and how they influence the outcome of aggregation functions.

The result of the analysis concerning the node \lstinline+Th+ in
\figurename~\ref{fig:nodo-semplice}, is as follows.
We will discuss in more detail this example in Section~\ref{sec:rt}.
\begin{itemize}
\item $\widehat{\Sigma}$ predicts, among other facts, that
  \lstinline+x+ may (and actually will) store the temperature as computed by the thermometers
  and sent through the message \lstinline!(temp,mt)! mentioned above;
\item $\kappa$ contains the pair consisting of the termometer
  \lstinline+Th$_1$+ (the sender of the message) and an abstract
  value for the temperature, because of the input on
  line~\ref{line:rcvcont};
\item $\Theta$ predicts that the node \lstinline+Th+ may
  handle values 
  (those stored in \lstinline+x+)
  that are abstractly represented by a grammar that
  generates trees with unbounded depth.
\end{itemize}

%

%
%

%
The results of the analysis provide designers with the basis for
detecting possible flaws on the usage of data as early as possible during the design phase.
They can also be used for checking and certifying various properties of IoT systems.
For example, one can inspect the component $\kappa$ for verifying if
the value of a certain sensor reaches a specific smart object in the
network and, with the component $\Theta$, if it is used as expected.
This kind of checks will be shown in Section~\ref{sec:rt} on the system in \figurename~\ref{fig:versione1}.
One can see there that the
value of the temperature of \lstinline+Th$_1$+ is indeed
received by the component \lstinline+Th$_2$+, but never used,
suggesting that the specification in hand is flawed.
In addition, a designer can check if the system conforms to the security requirements in force, so \IoTLySa\ supports a \emph{security by design} development mode.
For example, the CFA is used in~\cite{BDFG_ICE2016} to verify secrecy and a classical access control policy.
Indeed, one can inspect the component $\kappa$ and see if a secret value flows in clear on the network.
Other policies are also checked in~\cite{BodeiDFG16a}.

In order to experiment on the CFA, we had to adapt the one in~\cite{BDFG_Coord16,BodeiDFG16a},
so as to take into account the new constructs for the external choice mentioned above.
In particular, our extensions help in establishing a finer relation between the program points where a message is sent and where it is received. 
The implementation of the constraint solver for this improved CFA is still in an early stage of development,
and is not described here. 
%
%


\section{CFA at Work}\label{sec:rt}
We now describe our simple scenario and show how to analyse it using the CFA described in Section~\ref{sec:iotlysa}.
Suppose you have to design an IoT system for keeping the temperature
of a room under control through an air conditioner.
The system, dubbed \sysname\ (after AC controller), consists of three
nodes each specifying a \quo{smart} object: the thermostat discussed
in Section~\ref{sec:bkg}, and two thermometers, dubbed
\lstinline+Th$_1$+ and \lstinline+Th$_2$+, which
communicate through a wireless network.
We assume that thermometers have different physical features:
\lstinline+Th$_1$+ works on batteries for the lack of sockets in
the corner of the room where it is positioned, while
\lstinline+Th$_2$+ is mains powered.

A first version of the specification is in
\figurename~{\ref{fig:versione1}, where we omit the thermostat
  displayed in \figurename~\ref{fig:nodo-semplice}.
\begin{figure}[t]
%
%
\begin{lysacode}
func quit       : int £\label{line:sigstart}£
func temp       : int
func on         : bool
func off        : bool
func tthreshold : float
func bthreshold : float £\label{line:sigend}£

/*  the enumerated type of actions that the actuator of the Th can perform: */
actuator_type onoff_t {turnon, turnoff} £\label{line:onoff}£

node Th$_1$ = 
  sensor Temp1 : float = rec h. tau; probe; h
  sensor Battery : float = rec h. tau; probe; h

  process =
    i := 0;      £\label{line:init}£
    rec h.
      if bthreshold < Battery then        £\label{line:bguard}£
        recv(temp;x);     £\label{line:rcvtemp}£
        t := Temp1;   £\label{line:tp}£
        tp := t + x;      
        mt := tp / 2; £\label{line:mt}£
        i := i + 1;     £\label{line:incr}£
        if i = 10 then
          snd (temp,mt) to [Th,Th$_2$];      £\label{line:a-entrambi}£
          recv(ack;j);     £\label{line:riceve-j}£
          i := 0;
          h
        else
          snd (temp,mt) to [Th$_2$];  £\label{line:solo-a-2}£
          h
      else
        snd (quit,0) to [Th]; £\label{line:tstart}£
        nil £\label{line:tstop}£

  process = £\label{line:thread2start}£
    rec h.
      recv(temp;s);
      mt := s;
      h £\label{line:thread2stop}£



node Th$_2$ = £\label{line:t2start}£
  sensor Temp2 : float = rec h. tau; probe; h

  process = 
    mt := 0;
    rec h.
      snd (temp,mt) to [Th$_1$];
      recv(temp;x); £\label{line:t2receive}£
      t := Temp2; £\label{line:t2assignstart}£
      tp := t;
      mt := tp / 2; £\label{line:t2assignstop}£
      h £\label{line:t2stop}£
\end{lysacode}
\caption{The specification of \sysname\ in \IoTLySa}
\label{fig:versione1}
\end{figure}
A specification defines, before the behaviour of
nodes, a \emph{signature} declaring the types, functions, and
constants used inside nodes.
The signature of our example is on
lines~\ref{line:sigstart}-\ref{line:sigend} of
\figurename~\ref{fig:versione1} and specifies the constants used in
our scenario.
The syntax of a constant definition is \lstinline+func <name>:<type>+
where \lstinline+func+ is a keyword and \lstinline+<type>+ can
be a primitive type as \lstinline+int+, \lstinline+bool+ and
\lstinline+float+ or user-defined abstract types.\footnote%
{
At the moment \IoTLySa\ supports the declaration of abstract types
through the syntax \lstinline+type <name>+.
A more powerful and general form of user-defined data types is left as
future work. }
 The first four constants are used
in the messages exchanged in the system, the last two represent the
temperature threshold checked by the thermostat \lstinline+Th+ in
\figurename~\ref{fig:nodo-semplice} and the battery threshold tested
by \lstinline+Th$_1$+ (line~\ref{line:bguard}).
This node has sensors \lstinline+Temp1+ and \lstinline+Battery+; the
first cyclically senses the environment through the \lstinline+probe+ operation, and
returns a value of type \lstinline+float+ representing the
temperature; similarly, the other sensor returns a \lstinline+float+
representing the battery level;
the actions \lstinline+tau+ occurring at lines 12, 13 and 45 denote internal activities not relevant here, e.g.\ noise reduction, value normalization, etc.
Sensors operate at certain rates not explicitly given in the
specification.

We first consider the behaviour of \lstinline+Th$_2$+, specified at
lines~\ref{line:t2start}-\ref{line:t2stop}.
%
The thermometer uses the sensor \lstinline+Temp2+ and repeatedly
\begin{itemize}
\item sends the current local temperature  \lstinline+mt+, initially set to zero;
\item receives the temperature (sent by \lstinline+Th$_1$+) in its local variable \lstinline+x+;
\item re-assigns \lstinline+mt+ with half of the current  temperature sensed by \lstinline+Temp2+.
\end{itemize}
Clearly, the last step is not what we have in mind, i.e.\ re-assigning \lstinline+mt+ with the \emph{average} of the current
temperature sensed by \lstinline+Temp2+ and \emph{the one} received from \lstinline+Th$_1$+ (see the fixed version at line~\ref{line:fixmediat} in \figurename~\ref{fig:versione2}).
We will see below how the CFA detects this simple error.

The behaviour of \lstinline+Th$_1$+ is more complex
because it also  interacts with \lstinline+Th+, provided that its battery has enough charge.
The node \lstinline+Th$_1$+ has two parallel processes that share the local store.
In particular they both share and use the variable \lstinline+mt+.
%
%
%
\mbox{The first process repeatedly }
\begin{itemize}
\item receives in \lstinline+x+ the temperature sent by
  \lstinline+Th$_2$+ (line~\ref{line:rcvtemp});
\item computes the average with the current recorded temperature
  (lines~\ref{line:tp}-\ref{line:mt});
\item 
either sends the computed temperature to \lstinline+Th$_2$+ only (line~\ref{line:solo-a-2}, when the number \lstinline+i+ of iterations is not 10) or it multicasts the temperature to both \lstinline+Th+ and \lstinline+Th$_2$+  (line~\ref{line:a-entrambi}, when $\mathsf{i} = 10$);
\item waits for an \lstinline+ack+,
\end{itemize}
provided that the battery power is over the threshold
\lstinline+bthreshold+.
Otherwise, \lstinline+Th$_1$+ stops the interactions with
\lstinline+Th+ (lines~\ref{line:tstart}-\ref{line:tstop}).
The second process
(lines~\ref{line:thread2start}-\ref{line:thread2stop}) simply pings
\lstinline+Th$_2$+, by continuously receiving and storing the
temperatures received from \lstinline+Th$_2$+.

The results of our first analysis of ACC concern 
the usage of data and are in \figurename~\ref{fig:cfa}.
As said above, aggregation functions occurring in a process can grow terms of unbound depth.
For example, the variable \lstinline+i+ of \lstinline+Th$_2$+ gets increasingly actual values, starting from 0, as the simple aggregation function + is applied to the value just computed and 1.
Abstractly, this would be represented by a tree of the form
\lstinline!+(1,+(1,...+(0,1)...)!\!\!, where we still represent the abstract function with \lstinline!+!, and the tree as the string resulting from a preorder visit.
To keep the representation of abstract values finite (and get rid of their unknown depth), we use here regular tree grammars, cf.~\cite{BodeiDFG16a}.
For saving space, the entries in \figurename~\ref{fig:cfa} are the start symbols of the relevant tree grammars, while their productions are in the lower part of the figure.
Back to our little example, from the expression \lstinline!i+1!, the analysis extracts the  following grammar, with start symbol $G$ and the two productions (actually we are omitting some superscripts, see below):
\[
G \quad \rightarrow \quad 0  \quad \mid  \quad +(G,1)
\]
The first production reflects the assignment at line~\ref{line:init}, the second one the occurrence of the aggregation function in the expression at line~\ref{line:incr}.
The abstract value $G$ is then one of the possible results of the map $\widehat{\Sigma}$ on \lstinline+i+, as far as the node \lstinline+Th$_1$+ is concerned.

%

As said in Section~\ref{sec:bkg}, the component $\kappa$ of the
analysis (over-)approximates the messages that a node can receive  from
other nodes.
For instance, in our running example
\begin{center}
  (\lstinline+Th+,
  \guillemotleft{ack$^{\mathsf{Th}}$,on$^{\mathsf{Th}}$}\guillemotright) $\in \kappa$(\lstinline+Th$_1$+)
\end{center}
results from the analysis of the code in
\figurename~\ref{fig:nodo-semplice} at line~\ref{line:sndon}, and
indicates that the thermostat node \lstinline+Th+ could send a message
$(\mathsf{ack}, \mathsf{on})$ to the first thermometer \lstinline+Th$_1$+.
The superscripts appearing in the abstract values record the provenance
(in this case that the values have been originated from \lstinline+Th+).
%
%
Note in passing that in the abstract message $\mess{\mathsf{ack}^{Th},\mathsf{on}^{Th}}$,
$\mathsf{ack}^{Th}$ and $\mathsf{on}^{Th}$ should rather be replaced by two non-terminals, acting as start symbols of
two tree grammars defining the single leaves \lstinline+ack+ and
\lstinline+on+, respectively (see also \figurename~\ref{fig:cfa}).
For the sake of simplicity, we prefer here to only write the single leaves they generate.
%

Furthermore, the abstract value $\mathsf{on}^{Th}$ is
propagated in the component $\widehat{\Sigma}$:
\[
  \mathsf{on}^{Th} \in \widehat{\Sigma}_{\mathsf{Th}_1}(\mathsf{j}) 
\]
This happens because the abstract message $\mess{\mathsf{ack}^{Th},\mathsf{on}^{Th}}$ is considered when computing the abstract value predicted to bind \lstinline+j+ occurring in \lstinline+recv(ack;j)+ at line~\ref{line:riceve-j}. 
This propagation reflects that the concrete value abstracted as $\mathsf{on}^{Th}$ can at run-time be stored by
\lstinline+Th$_1$+  in its local variable $\mathsf{j}$.
Also, \lstinline+Th$_1$+ accepts two further messages (the guards match) from \lstinline+Th$_2$+, and the analysis says that
\[
(\mathsf{Th}_2,  \mess{\mathsf{temp}^{\mathsf{Th}_2}, 0^{\mathsf{Th}_2}}), (\mathsf{Th}_2,  \mess{\mathsf{temp}^{\mathsf{Th}_2}, D^{\mathsf{Th}_2}}) \in \kappa(\mathsf{Th}_1)
\]
The presence of $0^{\mathsf{Th}_2}$ reflects that in the first iteration of
\lstinline+Th$_2$+ sends $0$, the contents of $\mathsf{mt}$.
The second abstract value is $D^{\mathsf{Th}_2}$, the start symbol of the
following tree grammar:
\[
D^{\mathsf{Th}_2}  \quad \rightarrow  \quad /(\mathsf{Temp2}^{\mathsf{Th}_2}, 2^{\mathsf{Th}_2})
\]
This grammar accounts for the chain of assignments at
lines~\ref{line:t2assignstart}-\ref{line:t2assignstop} and, together
with $\widehat{\Sigma}_{\mathsf{Th}_2}$, establishes that
\lstinline+Th$_2$+ stores in \lstinline+mt+ half of the actual
value sensed by \lstinline+Temp2+.
Now, these abstract values populate
$\widehat{\Sigma}_{\mathsf{Th}_1}(\mathsf{x})$, and also flow in
$ \widehat{\Sigma}_{\mathsf{Th}_1}(\mathsf{mt})$, because the CFA mimics
the execution of lines~\ref{line:tp}-\ref{line:incr}.
In addition, the expressions in lines~\ref{line:tp} and~\ref{line:mt}
originate the grammars
\[
F^{\mathsf{Th}_1}  \quad \rightarrow  \quad +(\mathsf{Temp1}^{\mathsf{Th}_1}, 0^{\mathsf{Th}_2})  \quad \mid  \quad +(\mathsf{Temp1}^{\mathsf{Th}_1},   D^{\mathsf{Th}_2} )
\qquad\text{and}\qquad
D^{\mathsf{Th}_1}  \quad \rightarrow  \quad /(F^{\mathsf{Th}_1}, \mathsf{2}^{\mathsf{Th}_1})
\]
Therefore,  $D^{\mathsf{Th}_1} \in \widehat{\Sigma}_{\mathsf{Th}_1}$(\lstinline!mt!), which is then propagated in $\Theta(\mathsf{Th}_1)$.
Note that its language includes the abstract value \lstinline=/(+(Temp1,/(Temp2,2)),2)=, predicting that the actual value of \lstinline+mt+ occurring in \lstinline+Th$_1$+ may also depend on data computed by
\lstinline+Th$_2$+: this further possibility is recorded by the inclusion $D^{\mathsf{Th}_2} \in \widehat{\Sigma}_{\mathsf{Th}_1}$(\lstinline!mt!).

Turning our attention to \lstinline+Th$_2$+, we unveil a problem.
The analysis correctly predicts
that \lstinline+Th$_2$+ may receive the temperature from \lstinline+Th$_1$+
because $(\mathsf{Th}_1, \mess{\mathsf{temp}^{\mathsf{Th}_1}, D^{\mathsf{Th}_1}}) \in \kappa(\mathsf{Th}_2)$.
Nonetheless, the CFA shows
that average of temperatures in the variable \lstinline+mt+ of
\lstinline+Th$_2$+ does not depend on the values received from
\lstinline+Th$_1$+, as we would like to be.
While the analysis of \lstinline!recv(temp;x)! at line~\ref{line:t2receive} causes
\[
D^{\mathsf{Th}_1} \in \widehat{\Sigma}_{\mathsf{Th}_2}\text{(\lstinline+x+)} 
\]
\figurename~\ref{fig:cfa} shows that instead
$D^{\mathsf{Th}_1}$ does not occur in $\widehat{\Sigma}_{\mathsf{Th}_2}$(\lstinline!mt!).
Since our analysis is sound, no abstract values of \lstinline+Th$_1$+ affect the average temperature computed by \lstinline+Th$_2$+.
Indeed, the value of $\mathsf{mt}$ does not depend on ${\mathsf x}$.
The flaw anticipated above is therefore detected.

A (very) simple inspection of the code suggests the designer the obvious fix to \lstinline+Th$_2$+, depicted in \figurename~\ref{fig:versione2}.
Analysing this version in particular causes the following modifications to the results of \figurename~\ref{fig:cfa}:
\begin{itemize}
\item adding $D^{\mathsf{Th}_1} \in \widehat{\Sigma}_{\mathsf{Th}_2}$(\lstinline+mt+) and $F^{\mathsf{Th}_2} \in \Theta(\mathsf{Th}_2)$; 
\item substituting  $F^{\mathsf{Th}_2}$ for $\widehat{\Sigma}_{\mathsf{Th}_2}$(\lstinline+tp+), where the tree grammar for  $F^{\mathsf{Th}_2}$  has the following productions:
\[
F^{\mathsf{Th}_2}  \quad \rightarrow  \quad
              +(\mathsf{Temp2}^{\mathsf{Th}_2}, 0^{\mathsf{Th}_2}) \quad \mid \quad +(\mathsf{Temp2}^{\mathsf{Th}_2},  D^{\mathsf{Th}_1})  \quad \mid  \quad +(\mathsf{Temp2}^{\mathsf{Th}_2},  D^{\mathsf{Th}_2})
\]
\end{itemize}
Now the mutual interdependence of the average temperatures computed by the two nodes is made evident by the mutual recursion of the productions for $D^{\mathsf{Th}_1}$ and $D^{\mathsf{Th}_2}$ through the nonterminals $F^{\mathsf{Th}_1}$ and $F^{\mathsf{Th}_2}$.
Note also that the language of $D^{\mathsf{Th}_2}$ accounts for the case when \lstinline+Th$_1$+ only pings \lstinline+Th$_2$+, through the last production $F^{\mathsf{Th}_2} \rightarrow +(\mathsf{Temp2}^{\mathsf{Th}_2},  D^{\mathsf{Th}_2})$ above (the superscripts make it clear that in this case the average is computed on the values of \lstinline+Th$_2$+, only).

\begin{figure}[t]
\footnotesize{

\noindent

\begin{tabular}{l|l|}
  \multicolumn 2 l {$\ \ \kappa$}
  \\
  \hline
  $\mathsf{Th}_1$ & $\{ (\mathsf{Th}, \mess{\mathsf{ack}^{\mathsf{Th}},\mathsf{on}^{\mathsf{Th}}}), (\mathsf{Th}, \mess{\mathsf{ack}^{\mathsf{Th}},\mathsf{off}^{\mathsf{Th}}}), (\mathsf{Th}_2, \mess{\mathsf{temp}^{\mathsf{Th}_2}, \mathsf{0}^{\mathsf{Th}_2}}),(\mathsf{Th}_2,  \mess{\mathsf{temp}^{\mathsf{Th}_2}, D^{\mathsf{Th}_2}}) \}$ 
  \\
  \hline
  $\mathsf{Th}_2$ & $\{ (\mathsf{Th}_1, \mess{\mathsf{temp}^{\mathsf{Th}_1}, D^{\mathsf{Th}_1}}), (\mathsf{Th}_1, \mess{\mathsf{temp}^{\mathsf{Th}_1}, D^{\mathsf{Th}_2}}), (\mathsf{Th}_1, \mess{\mathsf{temp}^{\mathsf{Th}_1}, \mathsf{0}^{\mathsf{Th}_2}}) \}$  \\
  \hline
  $\mathsf{Th}$ & $\{ (\mathsf{Th}_1, \mess{\mathsf{temp}^{\mathsf{Th}_1}, D^{\mathsf{Th}_1}}), (\mathsf{Th}_1, \mess{\mathsf{temp}^{\mathsf{Th}_1}, D^{\mathsf{Th}_2}}), (\mathsf{Th}_1, \mess{\mathsf{temp}^{\mathsf{Th}_1}, \mathsf{0}^{\mathsf{Th}_2}}), (\mathsf{Th}_1, \mess{\mathsf{quit}^{\mathsf{Th}_1}, \mathsf{0}^{\mathsf{Th}_1}}) \}$  \\
  \hline
\end{tabular}
\\
\medskip

\begin{tabular}{l|l|}
  \multicolumn 2 l {$\ \ \Theta$}
  \\
  \hline
  $\mathsf{Th}_1$ & $\{ D^{\mathsf{Th}_1}, D^{\mathsf{Th}_2}, \mathsf{bthreshold}^{\mathsf{Th}_1}, \mathsf{Battery}^{\mathsf{Th}_1} , \mathsf{0}^{\mathsf{Th}_1}, G^{\mathsf{Th}_1}, \mathsf{temp}^{\mathsf{Th}_1},\mathsf{Temp1}^{\mathsf{Th}_1}, F^{\mathsf{Th}_1}, \mathsf{on}^{\mathsf{Th}}, \mathsf{off}^{\mathsf{Th}}, 0^{\mathsf{Th}_2}, $ \\
                  & $\ \  1^{\mathsf{Th}_1}, 2^{\mathsf{Th}_1}, 10^{\mathsf{Th}_1}  \}$ 
  \\
  \hline
  $\mathsf{Th}_2$ & $\{ D^{\mathsf{Th}_2}, \mathsf{temp}^{\mathsf{Th}_2}, \mathsf{Temp2}^{\mathsf{Th}_2}, 0^{\mathsf{Th}_2},2^{\mathsf{Th}_2}  \}$  \\
  \hline
  $\mathsf{Th}$ & $\{ D^{\mathsf{Th}_1}, D^{\mathsf{Th}_2} , 0^{\mathsf{Th}_2}, 0^{\mathsf{Th}_1}, \mathsf{ack}^{\mathsf{Th}}, \mathsf{on}^{\mathsf{Th}}, \mathsf{off}^{\mathsf{Th}},  \mathsf{quit}^{\mathsf{Th}}, \mathsf{tthreshold}^{\mathsf{Th}}  \}$  \\
  \hline
\end{tabular}
\medskip

\begin{multicols}{3}
%
  \begin{minipage}[c]{1.0\columnwidth}
    \begin{tabular}{l|l|}
      \multicolumn 2 l {$\ \ \widehat{\Sigma}_{\mathsf{Th}}$}\\
      \hline
      $\mathsf{x}$ & $\{D^{\mathsf{Th}_1}, D^{\mathsf{Th}_2} , \mathsf{0}^{\mathsf{Th}_2} \}$ 
      \\
      \hline
      $\mathsf{i}$ & $\{ \mathsf{0}^{\mathsf{Th}_1} \}$  \\
      \hline
    \end{tabular}
    \\[1em]
    \begin{tabular}{l|l|}
      \multicolumn 2 l {$\ \ \widehat{\Sigma}_{\mathsf{Th}_2}$}
      \\
      \hline
      $\mathsf{mt}$ & $\{ \mathsf{0}^{\mathsf{Th}_2}, D^{\mathsf{Th}_2} \}$ \\
      \hline
      $\mathsf{t}$ & $\{ \mathsf{Temp2}^{\mathsf{Th}_2}  \}$  \\
      \hline
      $\mathsf{tp}$ & $\{ \mathsf{Temp2}^{\mathsf{Th}_2}  \}$  \\
      \hline
      $\mathsf{x}$ & $\{ D^{\mathsf{Th}_1}, D^{\mathsf{Th}_2}, \mathsf{0}^{\mathsf{Th}_2}  \}$  \\
      \hline
      $\mathsf{Temp2}$ & $\{ \mathsf{Temp2}^{\mathsf{Th}_2}  \}$  \\
      \hline
    \end{tabular}
    \end{minipage}
  \columnbreak
  \begin{minipage}[c]{1.2\columnwidth}
    \begin{tabular}{l|l|}
      \multicolumn 2 l {$\ \ \widehat{\Sigma}_{\mathsf{Th}_1}$}
      \\
      \hline
      $\mathsf{i}$ & $\{ \mathsf{0}^{\mathsf{Th}_1}, G^{\mathsf{Th}_1}  \}$ \\
      \hline
      $\mathsf{j}$ & $\{\mathsf{on}^{\mathsf{Th}},\mathsf{off}^{\mathsf{Th}} \}$ \\
      \hline
      $\mathsf{mt}$ & $\{ D^{\mathsf{Th}_1}, D^{\mathsf{Th}_2} , \mathsf{0}^{\mathsf{Th}_2}  \}$ \\
      \hline
      $\mathsf{t}$ & $\{ \mathsf{Temp1}^{\mathsf{Th}_1} \}$  \\
      \hline
      $\mathsf{tp}$ & $\{ F^{\mathsf{Th}_1} \}$  \\
      \hline
      $\mathsf{x}$ & $\{ \mathsf{0}^{\mathsf{Th}_2}, D^{\mathsf{Th}_2}  \}$  \\
      \hline
      $\mathsf{s}$ & $\{ \mathsf{0}^{\mathsf{Th}_2}, D^{\mathsf{Th}_2} \}$  \\
      \hline
      $\mathsf{Battery}$ & $\{ \mathsf{Battery}^{\mathsf{Th}_1}  \}$ \\
      \hline
      $\mathsf{Temp1}$ & $\{ \mathsf{Temp1}^{\mathsf{Th}_1} \}$  \\
      \hline
    \end{tabular}
  \end{minipage}
 \begin{minipage}[c]{1.0\columnwidth}
   $ \begin{array}{lll}
   \\
   \\
   \\
      G^{\mathsf{Th}_1} & \rightarrow & \mathsf{0}^{\mathsf{Th}_1} \,\, \mid \,\, +(G^{\mathsf{Th}_1},\mathsf{1}^{\mathsf{Th}_1})
      \\
      D^{\mathsf{Th}_1} & \rightarrow &  /(F^{\mathsf{Th}_1}, \mathsf{2}^{\mathsf{Th}_1}) 
\\
F^{\mathsf{Th}_1} & \rightarrow &  +(\mathsf{Temp1}^{\mathsf{Th}_1}, 0^{\mathsf{Th}_2})
\\
& \mid &
 +(\mathsf{Temp1}^{\mathsf{Th}_1},   D^{\mathsf{Th}_2} )
\\
D^{\mathsf{Th}_2} & \rightarrow &  /(\mathsf{Temp2}^{\mathsf{Th}_2}, \mathsf{2}^{\mathsf{Th}_2}) 
          \end{array}$
\end{minipage}
\end{multicols}
}

\bigskip

\noindent
where 
%
by abuse of notation, we only write the leaves instead of writing a production, e.g.\ $0^{\mathsf{Th}_1}$ in place of $Zero^{\mathsf{Th}_1} \rightarrow 0^{\mathsf{Th}_1}$.
\caption{The results of the CFA for the system in \figurename~\ref{fig:versione1}}
\label{fig:cfa}
\end{figure}

\begin{figure}[b]
\begin{lysacode}
node Th$_2$ = 
  sensor Temp2 : float = rec h. tau; probe; h

  process = 
    mt := 0;
    rec h.
      snd (temp,mt) to [Th$_1$];
      recv(temp;x);
      t := Temp2;
      tp := t + x;  £\label{line:fixmediat}£
      mt := tp / 2;
      h
\end{lysacode}
\caption{Amended version of \lstinline+Th$_2$+}
\label{fig:versione2}
\end{figure}

\begin{figure}[b]
\begin{lysacode}
         process = 
         rec h.
             recv(temp;s);
             mt := s;
             snd(temp,mt) to [Th$_2$]; £\label{line:fixsend}£
             h
\end{lysacode}
\caption{A corrected  version of the second process of \lstinline+Th$_1$+}
\label{fig:versione3}
\end{figure}


\section{Bridging \IoTLySa\ and \chorgram}\label{sec:encoding}
%

The first step to bridge \IoTLySa\ with \chorgram\ is compiling 
a system specification into a set of CFSMs.
%
%
We give an informal account of the mapping; the technical definitions
are in~\cref{sec:bridge}; the tool implementing the translation is available on-line~\cite{iotlysatool}.

Each \IoTLySa\ node translates to a CFSM following the three steps
below:
\begin{enumerate}
\item \label{it:vars} replace each expression $e$ in
  \IoTLySa\ processes with a fresh variable $x_e$, and record the
  mapping $x_e \mapsto e$ in a specific environment;
\item \label{it:kappa} 
  extract from the component $\kappa$ of the analysis the abstract messages and simplify them so that they only contain tags and fresh variables as the ones computed above;
  
\item \label{it:aut} generate a machine for each node as follows
\begin{enumerate}
\item construct for each process of a node a communicating machine (we
  use a procedure similar to Thompson's construction~\cite{Aho86}, where we use
  the result of steps $(1)$ and $(2)$);
\item compute the parallel composition of the machines of the node in
  hand with the standard product algorithm for automata;
\item obtain the deterministic version of the product automaton computed in (b).
\end{enumerate}
\end{enumerate}

The communication soundness checked by \chorgram\ depends on the
sequence of the input/output actions of nodes regardless of the actual
data exchanged.
Thus, in step~\eqref{it:vars} we simplify the \IoTLySa\ nodes by
replacing each input/output tuple with a symbolic counterpart that
includes only choice tags and fresh variables.
Furthermore, we compute also a map that records the correspondences
between the fresh variables introduced and the expressions they
replace.
Also, in step~\eqref{it:kappa} we obtain an auxiliary data structure
that maps each node to the set of possible messages it may receive
together with the possible messages.
The compilation phase in step~\eqref{it:aut} works on the simplified
\IoTLySa\ specification just obtained.
%
This phase uses a mix of standard algorithms on automata (steps
\eqref{it:aut}b and \eqref{it:aut}c) and a customised version of
Thompson's algorithm that translates an \IoTLySa\ process into a
non-deterministic finite automaton (NFA).
%
%
Our construction is defined by induction on the syntax and relies on
automata with a distinguished state, dubbed the \emph{terminal} state,
used to combine automata together.
We sketch the definition of the construction for each syntactic case:
\begin{itemize}
\item \lstinline+nil+ is mapped on an automaton with a single state (both
  initial and terminal) and no transitions;
\item a conditional construct results in the union of the
  automata of the two branches (as usual we add a new initial with
  $\epsilon$-transition to the initial states of the automata of the
  branches automata and a new terminal state with
  $\epsilon$-transition from the terminal states of the automata of
  the branches);
   
\item iteration is translated as an automaton having a new
  initial state $q$ with an $\epsilon$-transition, to the initial state
  of the automaton for the body of the iteration, say $B$, and an
  $\epsilon$-transitions from the terminal state of $B$ to $q$ (to
  realise the loop);
\item the assignments and commands (to actuator) prefixes,
  require to first generate the CFSM, say $M$, for the continuation of the
  prefix and then a new initial state $q$ for the resulting automaton
  with an $\epsilon$-transition to the initial state of $M$;
\item the receive construct is translated following the same schema
  above: we generate a new state and link it to the automaton
  generated for the continuation; we add as many transition as the
  possible senders, each of which are labelled with a string having
  the form $\ain[@][@][][{\msg[(d1,\ldots,dr)]}]$ where $\q$ is the
  current node, $\p$ is a possible sender (computed in
  step~\eqref{it:kappa}), and the tuple $\mathsf{(d1,\ldots,dr)}$ is
  the message sent;
\item for the send construct \lstinline+snd(d1,...,dr) to+
  $\mathsf{[\q_1,\ldots,\q_m]}$ we add new states $q_0,\ldots,q_m$ to
  the states of the CFSM of the continuation (say $M$), we set $q_0$
  to be the initial state of the resulting machine, and we add an
  $\epsilon$-transition from $q_m$ to the initial state of $M$ as well
  as, for each state $0 \leq i \leq m$, a transition from $q_i$ to
  $q_{i+1}$ labelled by $\aout[@][B_i][][{\msg[(d1,...,dr)]}]$ where
  $\p$ is the name of the current node;
\item the \lstinline+switch+ construct results in the union of the
  automata of its branches, but the transition are labelled with
  strings of the form \lstinline+l0sender?(d1,...,dr)+ where
  \lstinline+l0+ is the current node and \lstinline+sender+ is a
  possible sender (computed in step $(2)$).
\end{itemize}
Then, we compute the product of the automata of processes and
make the overall result deterministic.
 \figurename~\ref{fig:3automi} shows the three
automata obtained for the system ACC of Section~\ref{sec:rt}.
%
%
Also, note that during the compilation we order the outputs of the multicast;
this is done for obtaining smaller automata and it is
faithful to the original semantics of \IoTLySa\ since CFSMs are asynchronous.
Also, messages on the transitions use the fresh variables
generated during the construction described above
(steps~\eqref{it:vars} and~\eqref{it:kappa}).
For example, the message of the left-most transition of
\lstinline+Th+ is $\mathsf{\conf{quit,m4}}$ where according to
the correspondence computed in the encoding $\mathsf{m4}$ is mapped to
the constant $\mathsf{0}$ which is sent by \lstinline+Th$_1$+ at
line~\ref{line:tstart} of \figurename~\ref{fig:versione1}.

Note that our compilation
preserves the communication pattern of \IoTLySa\ specification. 
Actually,  the encoding of a single multicast communication 
\lstinline+snd(msg) to+ $[\ell_1, ..., \ell_m]$
introduces no communications besides those to the nodes $\ell_1, ..., \ell_m$, in spite of linearisation, as CFSMs behave asynchronously. 
In addition, although conditional statements are compiled into non-deterministic choices as usually done in the field of static analysis,
%
%
this abstraction step does not affect our results.
The obtained CFSMs therefore soundly over-approximate the communication behaviour of the given \IoTLySa\ specification.
Summing up, if the CFSMs enjoy a given property, e.g. deadlock-freedom, so does the original \IoTLySa\ system.


\section{Analysing the Interactions}\label{sec:comm} 
%

\begin{figure}
  \centering
  \begin{multicols}{2}
   \!\!\!\!\!
    \includegraphics[scale=.25]{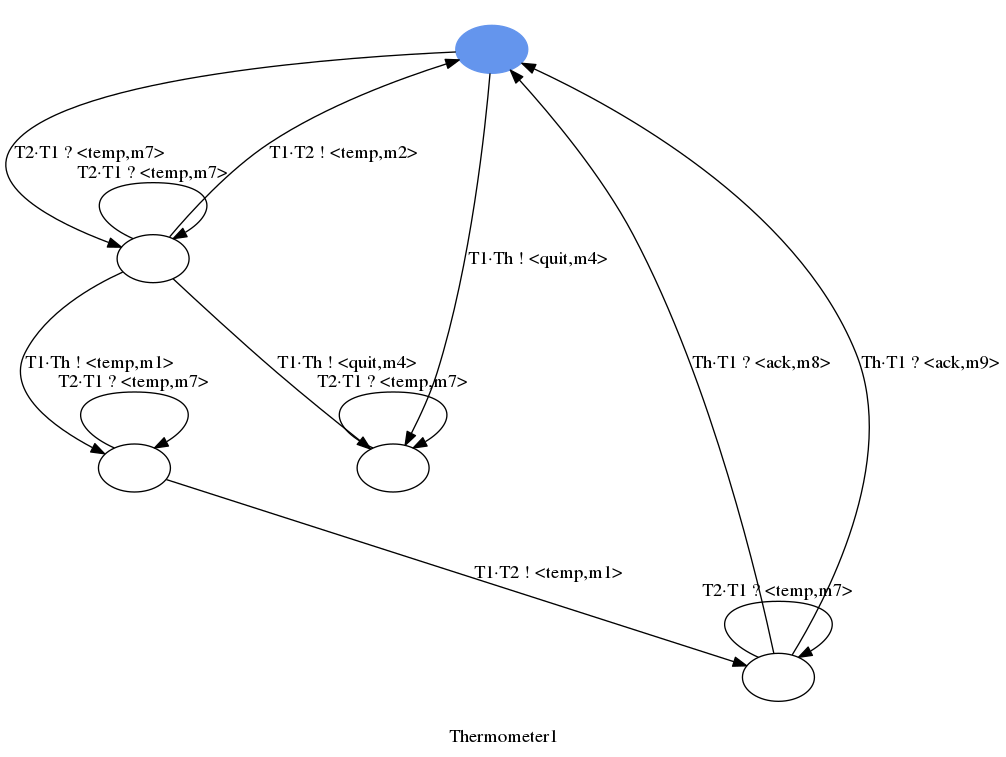}
    \columnbreak
    \begin{minipage}[c]{\columnwidth}
  \ \ \   \ \ \  \includegraphics[scale=.25]{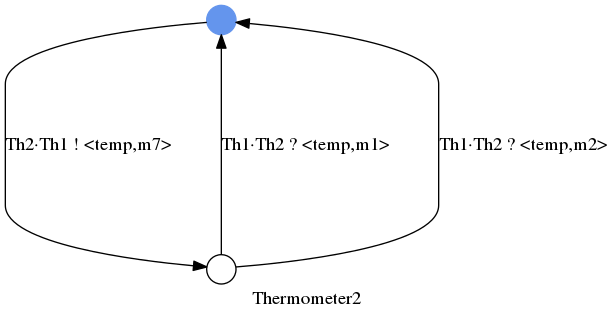}
    \\[2em]
   \ \ \   \ \ \    \includegraphics[scale=.25]{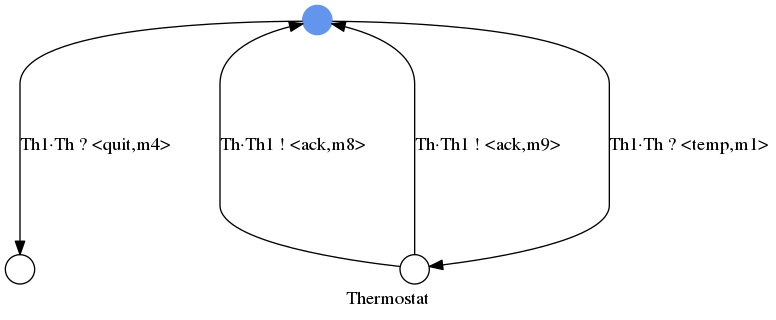}
  \end{minipage}
  \end{multicols}
  \caption{CFSMs of the amended specification \sysname}
  \label{fig:3automi}
\end{figure}

Our next step is to verify communication, in particular, to detect the presence of dangling or missing
messages left and of deadlocks.
%
%
To do that, we use \chorgram\ to check if the system enjoys the
generalised multiparty compatibility (GMC) mentioned in
Section~\ref{sec:cho} and described later.
To check GMC, \chorgram\ analyses the CFSMs of the system against
their synchronous executions, that is the finite labelled transition
system (STS for short) obtained by executing the system with two
additional constraints: (i) a message could be sent only if the
receiving partner is ready to consume the message and (ii) all buffers
are empty.

A system of CFSMs enjoys the GMC property when it is
\emph{representable} and has the \emph{branching
  property}~\cite{lty15}.
The representability condition essentially requires that for each
participant, the projection of the STS onto that participant yields a
CFSM bisimilar to the original machine.
The branching property condition requires that whenever a branching
occurs in the STS then either ($i$) the branching commutes, i.e.\ it
corresponds to two independent (concurrent) interactions, or ($ii$) it
corresponds to a choice that meets the following constraints:
\begin{enumerate}
\item
  \label{en:one-selector}
  there is a single participant, dubbed the \emph{selector}, that
  \quo{decides} which branch to take, and
\item
  \label{en:choice-aware}
  any participant not acting as a selector either has the same behaviour
  in each branch or it behaves differently in any
  branch of the choice.
  %
\end{enumerate}
Item~\eqref{en:one-selector} guarantees that every choice is located
at exactly one participant (this is crucial since we are assuming
\emph{asynchronous} communications).
Item~\eqref{en:choice-aware} ensures that all the other participants
are either unaware of the choice or, if they are involved, they can
distinguish which branch is taken by the selector.
We recall that the GMC property ensures communication
soundness, i.e.\ that systems are deadlock-free and they are not
message oblivious.

Giving the machines for \sysname\ (cf. Section~\ref{sec:encoding}) as
input to \chorgram\ we obtain the following STS:
\begin{center}
  \includegraphics[scale=.25]{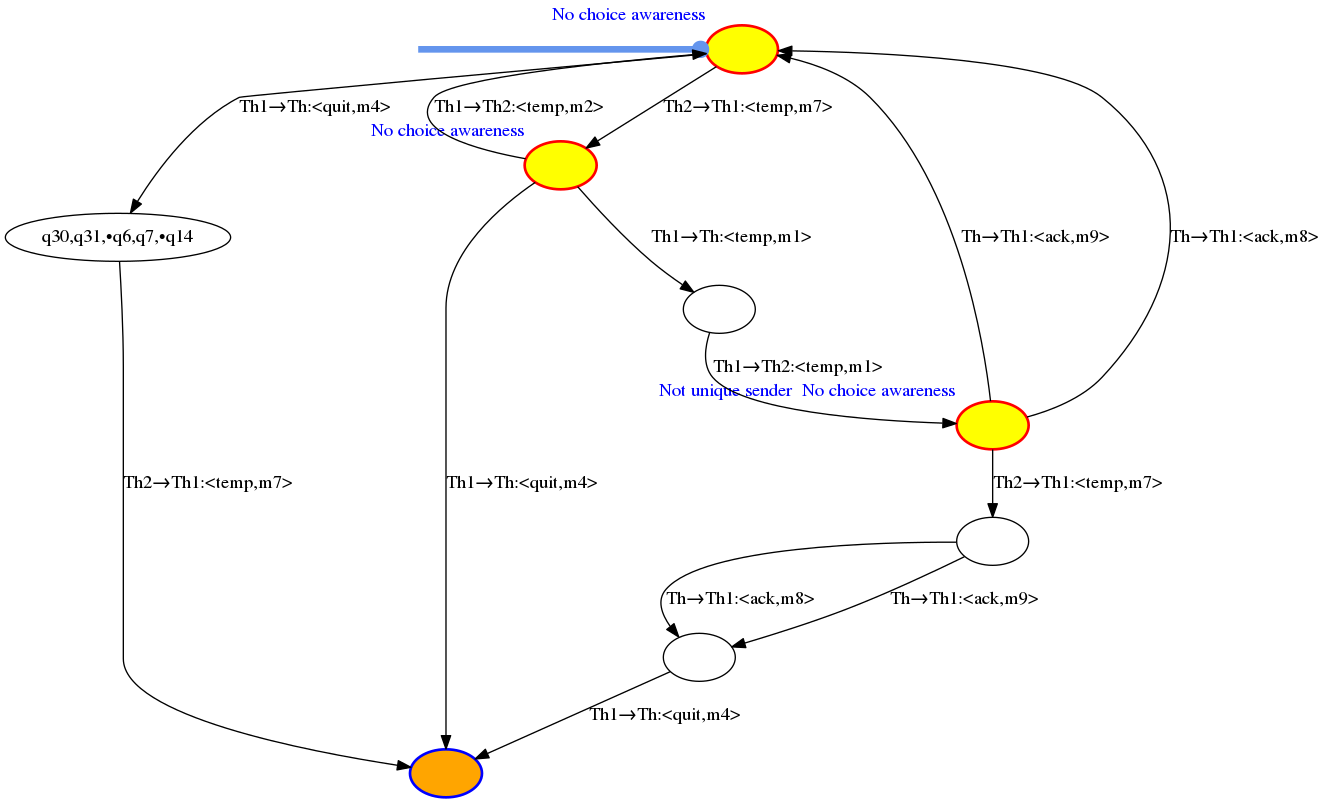}
\end{center}
where, for legibility, we adopt the abbreviations used in
Section~\ref{sec:rt} and omit the identities of irrelevant states.
The configurations of the STS are made of the states of the
(minimised) CFSMs of \sysname\ separated by a dot; for instance, the
left-most state in the STS above represents a configuration in which
the local state of (the CFSM of) \lstinline+Th$_1$+ is (the
equivalence class containing the) state $\mathsf{q30,q31}$,
\lstinline+Th$_2$+ is in state $\mathsf{q6,q7}$, and
\lstinline+Th+ is in state $\mathsf{q14}$.
This configuration, call it $c$, is reached when, from the initial
configuration, \lstinline+Th$_1$+ decides to send
\lstinline+Th+ the \lstinline+quit+ message, as described by
the left-most transition.
Note that \chorgram\ decorates this transition with the STS with the
\quo{No choice awareness} message to highlight that a potential
violation of the branching property occurs at the source state of the
transition (which is the initial state of the STS).
Indeed, configuration $c$ has a transition to a deadlock configuration,
also highlighted by \chorgram, hence detecting a communication
problem of \sysname.

To investigate and fix the problem we analyse the STS and the global
graph computed by \chorgram\ on the machines depicted in
\figurename~\ref{fig:3automi}.
\begin{figure}[h]
	\centering
	\includegraphics[width=\linewidth]{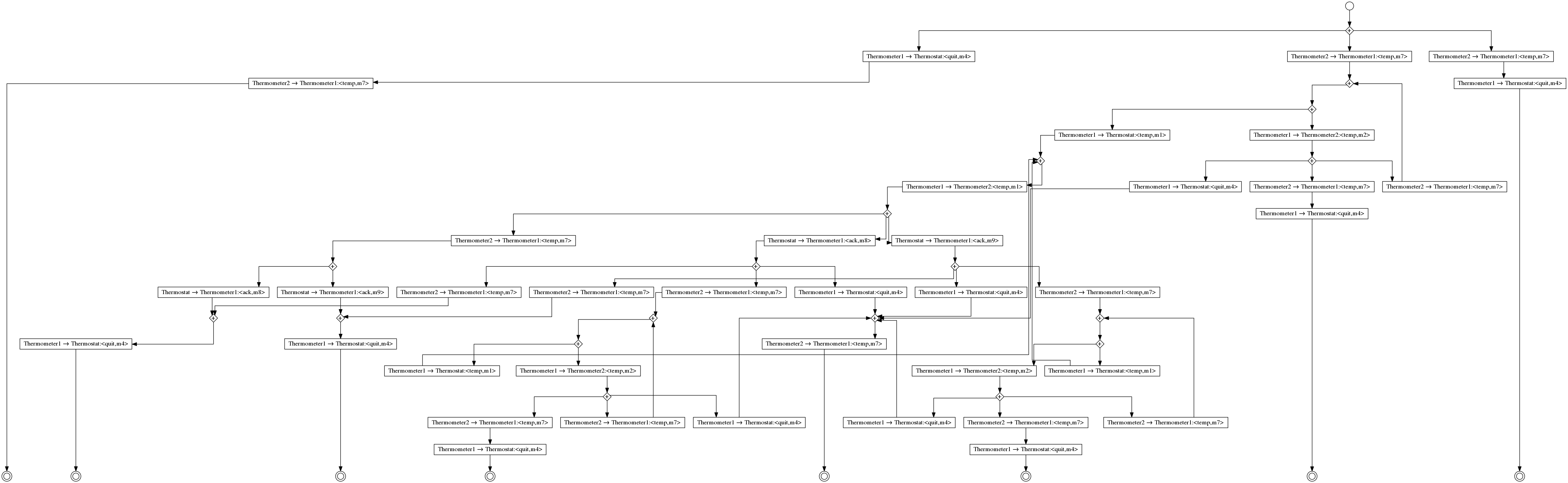}	
	\caption{The global graph of the communication-unsound \sysname}
	\label{fig:globalgraph1}
\end{figure}
Albeit not fully readable, we show the global graph in~\figurename~\ref{fig:globalgraph1} 
to give an idea of the complex
interactions that our simple scenario generates.
By inspecting this global graph we note that there are paths to sink nodes.
This is not expected of \sysname\ since each component is supposed to
perpetually execute its loop.
Therefore, those paths should be related to the reachable deadlock
configuration of the STS.
We start by analysing the left-most path ending in the sink node
\!\!\sinkG of \figurename~\ref{fig:globalgraph1} starting from the initial
node \!\sourceG in the graph.
This path describes a very short interaction only consisting of two
exchanged messages in which \lstinline+Th$_1$+ first sends
message \lstinline+quit+ to \lstinline+Th+ and then receives
the value of the temperature computed by \lstinline+Th$_2$+.
By inspecting the code in \figurename~\ref{fig:versione1}, the
attentive designer would observe that the first interaction
corresponds to the send performed by the first process of
\lstinline+Th$_1$+ at line \ref{line:tstart}.
After sending the \lstinline+quit+ message, the first process of
\lstinline+Th$_1$+ ends and makes the thermostat \lstinline+Th+ to
terminate as well (lines \ref{line:rcvquit}-\ref{line:end} of
\figurename~\ref{fig:nodo-semplice}).
Hence, the second interaction on the path under consideration can only
happen between \lstinline+Th$_2$+ and the second process of
\lstinline+Th$_1$+.
After sending its message \lstinline+Th$_2$+ waits for a reply
from \lstinline+Th$_1$+ (line~\ref{line:t2receive} of
\figurename~\ref{fig:versione1}); however, the second process of
\lstinline+Th$_1$+ sends no messages, it only waits for incoming
ones.
Thus, a deadlock occurs because \lstinline+Th$_1$+ and
\lstinline+Th$_2$+ are waiting each to receive from the other.

By inspecting the rest of the global graph, we discover that the other
deadlocks are caused by the same sequence of interactions: a
deadlock occurs as soon as \lstinline+Th$_1$+ sends the message
\lstinline+quit+ to the thermostat.
%
%
The problem arises because the second process of
\lstinline+Th$_1$+ never acknowledges the message received by
the other thermometer.
This is a violation of condition \eqref{en:choice-aware} of GMC (see
page~\pageref{en:choice-aware}).
As highlighted in the STS,
there is no choice awareness since the decision of quitting is not
properly propagated to all participants.
To fix this problem we change the code of the second process
\lstinline+Th$_1$+: we insert a send construct as shown in
\figurename~\ref{fig:versione3}, line \ref{line:fixsend}, so that the process replies to
each message of \lstinline+Th$_2$+.

Then, we re-apply \chorgram\ on the new version of our system and this
time \chorgram\ shows it is GMC, so establishing its communication
soundness.
Indeed, the extra communication from \lstinline+Th$_1$+ to
\lstinline+Th$_2$+ now ensures that the thermometer \lstinline+Th$_2$+ behaves
uniformly when communicating with \lstinline+Th$_1$+.
%
%
The combination of the two analyses is therefore
guaranteeing that our last version of \sysname\ does not mis-compute
or mis-communicate data.

%




\section{Conclusions}\label{sec:conc}
%

We proposed the combination of a control flow analysis and the
analysis of communication soundness based on a choreographic framework
as complementary approaches for the formal verification of IoT systems.
Both analyses and the interface between them are tool supported, although the constraint solver
for the first is still under development.
Our approach has been illustrated within a scenario that, albeit
conceptually simple, highlights non-trivial issues that designers of
IoT applications should tackle.
Indeed, tool support is crucial in this context since the high level
of non-determinism that IoT systems exhibit makes it is hard for the
designer to model robust applications.

For specifying IoT systems we appealed to \IoTLySa, a recently
proposed formalism based on a process calculus and tailored to
model monitoring system made of smart devices that communicate, 
make computations on sensed data, and command actuators.
%
%
It is endowed with a control flow analysis that allows the designer to identify points of the system
where computations are not as expected, both on the way data are aggregated and on the 
way communications occur among the smart objects.

Once the designer has attained a system correct from the point
of view of local computations,
communication properties are verified by checking that the components
of the IoT specification form a valid choreography using the
\chorgram\ tool.
  To link together the two analyses we implemented a compiler
  that translates a system specification into a set of CFSMs.
  The compiled system soundly reflects the communication behaviour of
  the original specification, ensuring that each property proved
  by \chorgram, also holds for the \IoTLySa\ specification.
During our
experiments, \chorgram\ has sometimes singled out unexpected
communications.
In particular, the complexity of the communication pattern of our
scenario was not fully understood before putting it under the lens of
\chorgram, even though our working examples were conceptually simple.

This paper is a first attempt to study the entanglement between local
computations and distributed interactions.
In our opinion, this issue is of uttermost relevance in IoT systems, as 
already discussed in Section~\ref{sec:intro}.
We are currently exploring how the combination of \IoTLySa\ and
its analysis with 
\chorgram\ could be used for verifying more sophisticated
properties of IoT systems.
In particular, we believe that the global model of an IoT system
returned by \chorgram\ could help us in refining the results of the control flow
analysis.
Indeed, the global model yields a more precise description of the
flow of data induced by actual communications than the current analysis of \IoTLySa\
can only approximate,
unless one makes it more complex and computationally expensive.
Improving the precision of the CFA facilitates detecting erroneous communications 
and misuse of data caused by them.
We further plan to investigate the definition of a modal logic for expressing and model checking a variety of properties
about communication and dependency of data on the global graph obtained by using our approach.



\bibliographystyle{eptcs}
\bibliography{biblio}

\begin{thebibliography}{10}
\providecommand{\bibitemdeclare}[2]{}
\providecommand{\surnamestart}{}
\providecommand{\surnameend}{}
\providecommand{\urlprefix}{Available at }
\providecommand{\url}[1]{\texttt{#1}}
\providecommand{\href}[2]{\texttt{#2}}
\providecommand{\urlalt}[2]{\href{#1}{#2}}
\providecommand{\doi}[1]{doi:\urlalt{http://dx.doi.org/#1}{#1}}
\providecommand{\bibinfo}[2]{#2}

\bibitemdeclare{misc}{bpmn}
\bibitem{bpmn}
\emph{\bibinfo{title}{{{B}usiness {P}rocess {M}odel and {N}otation}}}.
\newblock \bibinfo{note}{{\url{http://www.bpmn.org}}}.

\bibitemdeclare{misc}{iotlysatool}
\bibitem{iotlysatool}
\emph{\bibinfo{title}{Compiler from \IoTLySa\ to {CFSMs}}}.
\newblock \bibinfo{note}{\url{https://bitbucket.org/lillo/iotlysa}}.

\bibitemdeclare{book}{Aho86}
\bibitem{Aho86}
\bibinfo{author}{Alfred~V. \surnamestart Aho\surnameend}, \bibinfo{author}{Ravi
  \surnamestart Sethi\surnameend} \& \bibinfo{author}{Jeffrey~D. \surnamestart
  Ullman\surnameend} (\bibinfo{year}{1986}): \emph{\bibinfo{title}{Compilers:
  Principles, Techniques, and Tools}}.
\newblock \bibinfo{publisher}{Addison-Wesley}.

\bibitemdeclare{misc}{sch17}
\bibitem{sch17}
\bibinfo{author}{Amber \surnamestart Ankerholz\surnameend}
  (\bibinfo{year}{2017}): \emph{\bibinfo{title}{Bruce Schneier on New Security
  Threats from the Internet of Things}}.
\newblock
  \bibinfo{howpublished}{\url{https://www.linux.com/news/event/open-source-leadership-summit/2017/3/bruce-schneier-new-security-threats-internet-things}}.
\newblock \bibinfo{note}{Linux.com interviews Bruce Schneier}.

\bibitemdeclare{article}{bstz16}
\bibitem{bstz16}
\bibinfo{author}{Massimo \surnamestart Bartoletti\surnameend},
  \bibinfo{author}{Alceste \surnamestart Scalas\surnameend},
  \bibinfo{author}{Emilio \surnamestart Tuosto\surnameend} \&
  \bibinfo{author}{Roberto \surnamestart Zunino\surnameend}
  (\bibinfo{year}{2016}): \emph{\bibinfo{title}{Honesty by Typing}}.
\newblock {\sl \bibinfo{journal}{Logical Methods in Computer Science}}
  \bibinfo{volume}{12}(\bibinfo{number}{4}), \doi{10.2168/LMCS-12(4:7)2016}.

\bibitemdeclare{inproceedings}{BLY15}
\bibitem{BLY15}
\bibinfo{author}{Laura \surnamestart Bocchi\surnameend},
  \bibinfo{author}{Julien \surnamestart Lange\surnameend} \&
  \bibinfo{author}{Nobuko \surnamestart Yoshida\surnameend}
  (\bibinfo{year}{2015}): \emph{\bibinfo{title}{Meeting Deadlines Together}}.
\newblock In: {\sl \bibinfo{booktitle}{{CONCUR} 2015}}, pp.
  \bibinfo{pages}{283--296}, \doi{10.4230/LIPIcs.CONCUR.2015.283}.

\bibitemdeclare{article}{BBDNN_JCS}
\bibitem{BBDNN_JCS}
\bibinfo{author}{Chiara \surnamestart Bodei\surnameend},
  \bibinfo{author}{Mikael \surnamestart Buchholtz\surnameend},
  \bibinfo{author}{Pierpaolo \surnamestart Degano\surnameend},
  \bibinfo{author}{Flemming \surnamestart Nielson\surnameend} \&
  \bibinfo{author}{Hanne~Riis \surnamestart Nielson\surnameend}
  (\bibinfo{year}{2005}): \emph{\bibinfo{title}{Static validation of security
  protocols}}.
\newblock {\sl \bibinfo{journal}{Journal of Computer Security}}
  \bibinfo{volume}{13}(\bibinfo{number}{3}), pp. \bibinfo{pages}{347--390},
  \doi{10.3233/JCS-2005-13302}.

\bibitemdeclare{inproceedings}{BDFG_ICE2016}
\bibitem{BDFG_ICE2016}
\bibinfo{author}{Chiara \surnamestart Bodei\surnameend},
  \bibinfo{author}{Pierpaolo \surnamestart Degano\surnameend},
  \bibinfo{author}{Gian-Luigi \surnamestart Ferrari\surnameend} \&
  \bibinfo{author}{Letterio \surnamestart Galletta\surnameend}
  (\bibinfo{year}{2016}): \emph{\bibinfo{title}{A step towards checking
  security in {IoT}}}.
\newblock In: {\sl \bibinfo{booktitle}{Procs.~of ICE 2016}}, {\sl
  \bibinfo{series}{EPTCS}} \bibinfo{volume}{223}, pp.
  \bibinfo{pages}{128--142}, \doi{10.4204/EPTCS.223.9}.

\bibitemdeclare{inproceedings}{BDFG_Coord16}
\bibitem{BDFG_Coord16}
\bibinfo{author}{Chiara \surnamestart Bodei\surnameend},
  \bibinfo{author}{Pierpaolo \surnamestart Degano\surnameend},
  \bibinfo{author}{Gian-Luigi \surnamestart Ferrari\surnameend} \&
  \bibinfo{author}{Letterio \surnamestart Galletta\surnameend}
  (\bibinfo{year}{2016}): \emph{\bibinfo{title}{Where do your {IoT} ingredients
  come from?}}
\newblock In: {\sl \bibinfo{booktitle}{Procs.~of Coordination 2016}}, {\sl
  \bibinfo{series}{LNCS}} \bibinfo{volume}{9686},
  \bibinfo{publisher}{Springer}, pp. \bibinfo{pages}{35--50},
  \doi{10.1007/978-3-319-39519-7}.

\bibitemdeclare{article}{BodeiDFG16a}
\bibitem{BodeiDFG16a}
\bibinfo{author}{Chiara \surnamestart Bodei\surnameend},
  \bibinfo{author}{Pierpaolo \surnamestart Degano\surnameend},
  \bibinfo{author}{Gian-Luigi \surnamestart Ferrari\surnameend} \&
  \bibinfo{author}{Letterio \surnamestart Galletta\surnameend}
  (\bibinfo{year}{2017}): \emph{\bibinfo{title}{Tracing where IoT data are
  collected and aggregated}}.
\newblock {\sl \bibinfo{journal}{Logical Methods in Computer Science}}
  \bibinfo{volume}{13}(\bibinfo{number}{3:5}), pp. \bibinfo{pages}{1--38},
  \doi{10.23638/LMCS-13(3:5)2017}.

\bibitemdeclare{article}{bz83}
\bibitem{bz83}
\bibinfo{author}{Daniel \surnamestart Brand\surnameend} \&
  \bibinfo{author}{Pitro \surnamestart Zafiropulo\surnameend}
  (\bibinfo{year}{1983}): \emph{\bibinfo{title}{{On Communicating Finite-State
  Machines}}}.
\newblock {\sl \bibinfo{journal}{Journal of the ACM}}
  \bibinfo{volume}{30}(\bibinfo{number}{2}), pp. \bibinfo{pages}{323--342},
  \doi{10.1145/322374.322380}.

\bibitemdeclare{inproceedings}{dy12}
\bibitem{dy12}
\bibinfo{author}{Pierre{-}Malo \surnamestart Deni{\'{e}}lou\surnameend} \&
  \bibinfo{author}{Nobuko \surnamestart Yoshida\surnameend}
  (\bibinfo{year}{2012}): \emph{\bibinfo{title}{Multiparty Session Types Meet
  Communicating Automata}}.
\newblock In: {\sl \bibinfo{booktitle}{ESOP 2012}}, pp.
  \bibinfo{pages}{194--213}, \doi{10.1007/978-3-642-28869-2\_10}.

\bibitemdeclare{misc}{chorgram}
\bibitem{chorgram}
\bibinfo{author}{Julien \surnamestart Lange\surnameend} \&
  \bibinfo{author}{Emilio \surnamestart Tuosto\surnameend}
  (\bibinfo{year}{2015}): \emph{\bibinfo{title}{\chorgram: tool support for
  choreographic development}}.
\newblock \bibinfo{howpublished}{Available at {\chorgramsite}}.

\bibitemdeclare{inproceedings}{lty15}
\bibitem{lty15}
\bibinfo{author}{Julien \surnamestart Lange\surnameend},
  \bibinfo{author}{Emilio \surnamestart Tuosto\surnameend} \&
  \bibinfo{author}{Nobuko \surnamestart Yoshida\surnameend}
  (\bibinfo{year}{2015}): \emph{\bibinfo{title}{{From Communicating Machines to
  Graphical Choreographies}}}.
\newblock In: {\sl \bibinfo{booktitle}{POPL15}}, pp. \bibinfo{pages}{221--232},
  \doi{10.1145/2676726.2676964}.

\bibitemdeclare{inbook}{lty17}
\bibitem{lty17}
\bibinfo{author}{Julien \surnamestart Lange\surnameend},
  \bibinfo{author}{Emilio \surnamestart Tuosto\surnameend} \&
  \bibinfo{author}{Nobuko \surnamestart Yoshida\surnameend}
  (\bibinfo{year}{2017}): \emph{\bibinfo{title}{A tool for choreography-based
  analysis of message-passing software}}.
\newblock \bibinfo{note}{To appear. Available at
  \url{http://www.cs.le.ac.uk/~et52/chorgram_betty_ch.pdf}}.

\bibitemdeclare{article}{mil93turing}
\bibitem{mil93turing}
\bibinfo{author}{Robin \surnamestart Milner\surnameend} (\bibinfo{year}{1993}):
  \emph{\bibinfo{title}{{Elements of Interaction ­ Turing Award Lecture}}}.
\newblock {\sl \bibinfo{journal}{CACM}}
  \bibinfo{volume}{36}(\bibinfo{number}{1}), \doi{10.1145/151233.151240}.

\end{thebibliography}


\appendix

\section{\IoTLySa\ Syntax}\label{sec:applysa}
The syntax of systems
is as follows.

{\scriptsize
\[
\begin{array}{ll@{\hspace{2ex}}l} 
{\mathcal N} \ni N ::= & {\it systems \ of \ nodes} &\\
& \NIL                       & \hbox{empty system} \\
& \ell: [B]                      & \hbox{single node}\;  (\ell \in {\mathcal L})
\\
& N_1\ |\ N_2 &  \hbox{parallel composition of nodes}
\end{array}
\quad
\begin{array}{ll@{\hspace{2ex}}l}

{\mathcal B} \ni B ::= & \text {\it node components} &\\
&  \Sigma_\ell  & \hbox{store of node } \ell
\\
& P  & \hbox{process}
\\
& S   & \hbox{sensor, with a unique id $i \in \mathcal{I}_\ell$}
\\
& A     & \hbox{actuator, with a unique id $j \in \mathcal{J}_\ell$}
\\
& B \ \|\ B & \hbox{parallel composition of components}
\end{array}
\]
\[
\begin{array}{lll}
{\mathcal P} \ni P ::= & {\it control \ processes} &\\
& \NIL                       & \hbox{inactive process} \\
& \OUTM{E_1, \cdots, E_r}{L}.\,P & \hbox{asynchronous multi-output L$\subseteq {\mathcal L}$} \\
& \OUTM{c_1, \cdots,c_j, E_{j+1}\cdots E_r}{L}.\,P & \hbox{asynchronous multi-output (with tags)} \\
& \INPS{E_1,\cdots,E_j}{x_{j+1},\cdots,x_r}{P}\  & \hbox{input}\\
& \INPS{c_1,\cdots,c_j}{x_{j+1},\cdots,x_r}{P}  &  \hbox{input (with tags) }
\\
& \INPS{c_1,\cdots,c_j}{x_{j+1},\cdots,x_r}{P} + 
\\
&  \INPS{c'_1,\cdots,c'_j}{x'_{j+1},\cdots,x'_r}{Q} &  \hbox{switch or external choice }
\\
& E?P:Q &  \hbox{conditional statement} \\
& h   &  \hbox{iteration variable}
\\
&\mu h. \ P & \hbox{iteration}
  
\\[.2ex]
& x := E.\,P & \hbox{assignment to $x \in {\mathcal X}_\ell$}
\\
& \OUTS{j, \gamma}{P}& \hbox{output of action $\gamma$ to actuator $j$}
\end{array}
\quad
\begin{array}{ll@{\hspace{2ex}}l}
\\ \\
\hspace{-1cm}
{\mathcal E} \ni 
E ::= & 
       {\it terms}&\\
& v  & \hbox{value } (v \in {\mathcal V})\\
& i  & \hbox{sensor location } (i \in {\mathcal I_\ell})\\
& x  & \hbox{variable } (x \in {\mathcal X}_\ell)\\
& f(E_1, \cdots, E_r) &  \hbox{function on data }  (f \in {\mathcal F})\\
\end{array}
\]

\[
\begin{array}{ll@{\hspace{2ex}}l ll@{\hspace{2ex}}l}
S &::=  {\it sensors} & &
A &::= {\it actuators} &\\
& \NIL &  \hbox{\hspace{-1mm}inactive sensor} &&
 \NIL &  \hbox{\hspace{-1mm}inactive actuator}
\\
& \tau.S & \hbox{\hspace{-1mm}internal action} &
& \tau.A & \hbox{\hspace{-1mm}internal action} 
\\
& probe(i).\,{S} & \hbox{\hspace{-.8mm}sense a value by} 
&& 
(\!|j, \Gamma|\!).\,A & \hbox{\hspace{-1mm}command for actuator $j$} 
\\
&& \hbox{\hspace{-1mm}the $i^{th}$ sensor}
 &&
\gamma.A & \hbox{\hspace{-1mm}triggered action ($\gamma \in \Gamma$)}

\\
& h & \hbox{\hspace{-1mm}iteration variable} 
&&
h & \hbox{\hspace{-1mm}iteration variable} 
\\
& \mu h\,.\, S & \hbox{\hspace{-1mm}iteration}
&& \mu h\,.\, A & \hbox{\hspace{-1mm}iteration}
\end{array}
\]

}

We briefly comment on the output and input constructs.
The prefix $\OUTM{E_1, \cdots, E_r }{L}$ implements a simple multicast communication: the tuple $E_1, \dots, E_r$ is asynchronously sent to the nodes in $L$.
The input prefix $(E_1, \!\cdots\!,E_j; x_{j+1},\! \cdots \!,x_r)$
receives a $r$-tuple, provided that its first $j$ terms match the input ones, and then binds the remaining store variables (separated by a ``;'') to the corresponding values.
Otherwise, the $r$-tuple is not accepted.
The present variant of \IoTLySa\ has external
choices not originally included in~\cite{BDFG_Coord16,BDFG_ICE2016}\footnote{
  We recall that the \lstinline+switch+ construct can be encoded in
  the original presentation of \IoTLySa. } that is rendered by two branches guarded by
two different input actions.
Note that input and output actions may contain tags  $c \in \mathcal{C}$.

For instance the process included in the thermostat node \lstinline+Th+ in \figurename~\ref{fig:nodo-semplice}, is specified as follows.
\begin{align*}
P_{\mathsf{Th}} = \mu \mathsf{h}. & (\mathsf{temp};\mathsf{x}). (\mathsf{tthreshold} < \mathsf{x})\ ?\  \\
&  \qquad \OUTS{\mathsf{ACon\_off, on}}{}\OUTM{\mathsf{ack},\mathsf{on}}{\{\mathsf{Th_1}\}}\ .\ \mathsf{h} :\\
&  \qquad  \OUTS{\mathsf{ACon\_off, off}}{}\OUTM{\mathsf{ack},\mathsf{off}}{\{\mathsf{Th_1}\}}\ .\ \mathsf{h}\\
& \  + (\mathsf{quit};\mathsf{i}).\NIL
\end{align*}

\section{Mapping \IoTLySa\ to CFSMs}\label{sec:bridge}

Here, we detail the procedures described in Section~\ref{sec:encoding} for compiling an \IoTLySa\ specification into a set of CFSMs.
First, we define the simplification procedure that substitutes fresh variables for expressions in input/output prefixes.
This transformation is inductively defined on the syntax of the system of nodes:
\begin{definition}
Let  $N$ be an \IoTLySa\ system of nodes, the procedure $\mathit{simpl}\colon \mathcal{N} \to \mathcal{N}$
is inductively defined below, where we use two auxiliary functions $\mathit{simpl_B}\colon \mathcal{B} \to \mathcal{B}$ and 
$\mathit{simpl_P}\colon \mathcal{P} \to \mathcal{P}$.

{\small
\[
\begin{array}{ll}
\begin{array}{ll}
\mathit{simpl}(\NIL) &=\, \NIL \\
\mathit{simpl}(\ell: [B]) &=\, \ell: [\mathit{simpl_B}(B)] \\
\mathit{simpl}(N_1 \mid N_2) &=\, \mathit{simpl}(N_1) \mid \mathit{simpl}(N_2) 
\end{array}
&
\begin{array}{ll}
\mathit{simpl_B}(P) &=\, \mathit{simpl_P}(P) \\
\mathit{simpl_B}(B_1 \parallel B_2) &=\, \mathit{simpl_B}(B_1) \parallel \mathit{simpl_B}(B_2) \\
\mathit{simpl_B}(X) &=\, X \qquad \text{for } X \in \{\Sigma_\ell,\, S,\, A\} 
\end{array}
\end{array}
\]
\[
\begin{array}{ll}
\mathit{simpl_P}(X) &=\, X \\
                   & \qquad \text{for } X \in \{\NIL,\, h\}\\
\mathit{simpl_P}(Y.P) &=\, Y.\mathit{simpl_P}(P) \\
                   & \qquad \text{for } Y \in \{\mu h,\, x := E,\, \langle j, \gamma\rangle\} \\
\mathit{simpl_P}(E ? P_1 : P_2) &=\, E\, ?\, \mathit{simpl_P}(P_1) : \mathit{simpl_P}(P_2) \\                    
\mathit{simpl_P}(\OUTM{E_1, \cdots, E_r}{L}.\,P) &=\, \OUTM{m_1, \cdots, m_r}{L}.\,\mathit{simpl_P}(P)\\
                                                & \qquad m_1, \dots, m_r \text{ fresh variables}\\
\mathit{simpl_P}(\OUTM{c_1, \cdots, c_j, E_{j+1}, \cdots, E_r}{L}.\,P) &=\, \OUTM{c_1, \cdots, c_j, m_{j+1}, \cdots, m_r}{L}.\,\mathit{simpl_P}(P)\\
                                                & \qquad m_{j+1}, \dots, m_r \text{ fresh variables}\\
\mathit{simpl_P}(\INPS{E_1,\cdots,E_j}{x_{j+1},\cdots,x_r}{P}) &=\, \INPS{m_1,\cdots,m_j}{m_{j+1},\cdots,m_r}{\mathit{simpl_P}(P)} \\
& \qquad  m_1, \dots, m_r \text{ fresh variables}\\
\mathit{simpl_P}(\sum^2_{i=1} \INPS{c^i_1,\cdots,c^i_j}{x^i_{j+1},\cdots,x^i_r}{P_i}) &=\, \sum^2_{i=1} \INPS{c^i_1,\cdots,c^i_j}{m^i_{j+1},\cdots,m^i_r}{\mathit{simpl_P(P_i)}} \\
& \qquad m^i_{j+1}, \dots, m^i_r \,\text{ fresh variables}
\end{array}
\]
}
\label{def:simpl:proc}
\end{definition}

Applying the above procedure to the \IoTLySa\ process $P_{Th}$ above results in:
%
\begin{align*}
\mathit{simpl_P}(P_{\mathsf{Th}}) = \mu \mathsf{h}. & (\mathsf{temp};m_x). (\mathsf{tthreshold} < \mathsf{x})\ ?\  \\
&  \qquad \OUTS{\mathsf{ACon\_off, on}}{}\OUTM{\mathsf{ack},m_{\mathsf{on}}}{\{\mathsf{Th_1}\}}\ .\ \mathsf{h} :\\
&  \qquad  \OUTS{\mathsf{ACon\_off, off}}{}\OUTM{\mathsf{ack},m_{\mathsf{off}}}{\{\mathsf{Th_1}\}}\ .\ \mathsf{h}\\
& \  + (\mathsf{quit};m_{\mathsf{i}}).\NIL
\end{align*}

For brevity, we omitted above the straightforward computation of the environment.
Indeed, it sufficies modifying each procedure to return a pair whose second element is the environment;
to deal with recursive calls, e.g.\ in the case $E ? P_1 : P_2$, we combine the returned mappings to form the overall one.

\newcommand{\transA}[1]{\mathcal{A}\llbracket #1 \rrbracket \kappa}
Below, we formalise the compilation of a process inside an \IoTLySa\ node into an automaton.
This procedure operates on processes whose input/output prefixes only contain tags and fresh variables (cf.\ Definition~\ref{def:simpl:proc}).
We assume the standard definition of a finite state machine as a tuple $(S, s, F, \Sigma, \delta)$.
We define the function $\mathcal{A}\llbracket \_ \rrbracket \_ $ that takes a process, a component $\kappa$ of the CFA (modified in step~\ref{it:kappa} of Section~\ref{sec:encoding}, 
(e.g.\ $(\mathsf{Th},\mess{\mathsf{ack},m_{\mathsf{on}}}\in \kappa(\mathsf{Th_1})$ in $P_{\mathsf{Th}}$)
and returns a CFSM.
The obtained machine implements the same communication behaviour of the process.
For the sake of readability, in the following definition  we do not explicitly show the alphabet set of the automata
(simply the union of the alphabets of the automata returned by recursive calls and of the symbols (different from $\epsilon$) occurring in the transition relation $\delta$).
%
Furthermore, when needed we assume to generate new states, e.g.\ $s_i$ and $s_f$.
\begin{definition}
Let $P$ be an \IoTLySa\ process of the node $\ell$, and $\kappa$ be the CFA component (simplified in step~\ref{it:kappa} of Section~\ref{sec:encoding}). 
Then, the translation  
$\mathcal{A}\llbracket \_ \rrbracket \_ \colon \mathcal{P} \to \mathcal{K} \to CFSM$ is inductively defined as follows:
{\small
\[
\begin{array}{ll}
\transA{\NIL}      &=\, (\{s_i\},s_i, \emptyset,\emptyset)\\
\transA{h}         &=\, (\{s_i,\,s_f\},s_i, \{s_f\},\{(s_i, \epsilon, s_f)\})\\
\transA{Y.P}       &=\, (\{s_i\} \cup S^P,s_i, F^P,\delta^P \cup \{(s_i, \tau, s^P_i)\})\\
                   &\,\, \text{where } Y \in \{ x := E, \langle j, \gamma\rangle\}\,\, \text{ and } \transA{P} = (S^P,s^P_i, F^P,\delta^P)\\
\transA{E? P : Q}  &=\, (\{s_i,s_f\} \cup S^P \cup S^Q,s_i, \{s_f\},\delta^P \cup \delta^Q \cup \\ 
                   &\, \cup \{(s_i, \epsilon, s^P_i),(s_i, \epsilon, s^P_i)\} \cup \{(s, \epsilon, s_f) \mid s \in F^P \cup F^Q\})\\
                   &\,\, \text{where } \transA{P} = (S^P,s^P_i, F^P,\delta^P) \,\, \text{ and } \transA{Q} = (S^Q,s^Q_i, F^Q,\delta^Q)\\   
\transA{\OUTM{m_1, \cdots, m_r}{L}.\,P} &=\, (\{s_i,\,s_1,\,\dots,\,s_n\} \cup S^P, s_i,F^P, \delta^P \cup \bigcup^{n-1}_{i=0}(s_i,e_{i+1},s_{i+1}) \cup \{(s_n,\,\epsilon,s^P_i)\})\\
                   &\,\, \text{ where } \transA{P} = (S^P, s^P_i, F^P, \delta^P), \\
                   &\,\, s_1, \dots, s_n \text{ new, } L = \{\ell_1, \dots, \ell_n\}\\
                   &\,\, \text{ and } e_{i+1} = \ell\cdot \ell_{i+1} ! \mess{m_1,\dots, m_r}\\                   
\transA{\OUTM{m_1, \cdots, m_r}{L}.\,P} &=\, (\{s_i,\,s_1,\,\dots,\,s_n\} \cup S^P, s_i,F^P, \delta^P \cup \bigcup^{n-1}_{i=0}(s_i,e_{i+1},s_{i+1}) \cup \{(s_n,\,\epsilon,s^P_i)\})\\
                   &\,\, \text{ where } \transA{P} = (S^P, s^P_i, F^P, \delta^P), \\
                   &\,\, s_1, \dots, s_n \text{ new, } L = \{\ell_1, \dots, \ell_n\}\\
                   &\,\, \text{ and } e_{i+1} = \ell\cdot \ell_{i+1} ! \mess{m_1,\dots, m_r}\\
\transA{\OUTM{c_1, \cdots, c_j, m_{j+1}, \cdots, m_r}{L}.\,P} &=\, (\{s_i,\,s_1,\,\dots,\,s_n\} \cup S^P, s_i,F^P, \delta^P \cup \bigcup^{n-1}_{i=0}(s_i,e_{i+1},s_{i+1}) \cup \{(s_n,\,\epsilon,s^P_i)\})\\
                   &\,\, \text{ where } \transA{P} = (S^P, s^P_i, F^P, \delta^P), \\
                   &\,\, s_1, \dots, s_n \text{ new, } L = \{\ell_1, \dots, \ell_n\}\\
                   &\,\, \text{ and } e_{i+1} = \ell\cdot \ell_{i+1} ! \mess{c_1, \dots, c_j, m_{j+1}, \dots, m_r}\\
\transA{\INPS{m_1,\cdots,m_j}{m_{j+1},\cdots,m_r}{P}}  &=\, (\{s_i\} \cup S^P, s_i,F^P, \delta^P \cup \bigcup_{(\ell',t) \in \kappa(\ell)} (s_i,\ell\cdot\ell' ? t,s^P_i))\\
                   &\,\, \text{ where }  \transA{P} = (S^P,s^P_i,F^P,\delta^P) \\
\transA{\sum^2_{i=1} \INPS{c^i_1,\cdots,c^i_j}{m^i_{j+1},\cdots,m^i_r}{P_i}}    &=\,  
                     (\{s_i\} \cup \bigcup^{2}_{i=1}S^{P_i}, \bigcup^{2}_{i=1}F^{P_i}, \bigcup^{2}_{i=1}\delta^{P_i} \cup 
                               \bigcup_{(\ell',t) \in \kappa(\ell)}(s_i,\ell\cdot\ell' ? t, s^{P_i}_i)\\
                   &\,\, \text{ where }  \transA{P_i} = (S^{P_i}, s^{P_i}_i,F^{P_i},\delta^{P_i})\\                                         
\end{array}
\]
}
              
\end{definition}

Applying this translation to $P_{Th}$ results in the right-bottom automaton in \figurename~\ref{fig:3automi}.


\end{document}